\documentclass{iopjournal}
\usepackage{fix-cm}
\usepackage{amsmath,amssymb}
\usepackage{graphicx}
\usepackage{bm}
\usepackage{xcolor}
\makeatletter
\let\@makecaption\@undefined
\makeatother
\usepackage{subcaption}

\fancyhfoffset[L]{0pt}
\newcommand{\vect}[1]{\boldsymbol{\mathbf{#1}}}

\newcommand{\cut}[1]{{}}

\begin{document}

\title{Latent kinetic Ising models of neural spike trains}
\author{Davide Ghio\textsuperscript{*} and David Saad}\par
\affil{College of Engineering and Physical Sciences, Aston University, Birmingham, UK}\par
\vspace{8pt}
{\fontsize{8}{10}\selectfont\raggedright
\textsuperscript{*} Author to whom any correspondence should be addressed.\par}
\vspace{-10pt}
\email{davide.ghio98@gmail.com}

\begin{abstract}
Inferring directed effective interactions from neuronal spike trains is a central inverse problem in statistical physics and computational neuroscience. Kinetic Ising models provide a tractable framework for this task, but their application to neural data typically requires binning spike trains into binary activity variables, discarding within-bin timing and conflating collective network dynamics with single-neuron history effects. 

We introduce SpiKIsing, a latent-variable model that separates these two levels of description. A discrete-time asymmetric kinetic Ising model describes collective network activity, while continuous-time, history-dependent point processes generate the observed spikes conditional on the latent states, accounting explicitly for refractoriness and post-spike recovery. 

We derive a variational mean-field expectation-maximization scheme in which the point-process likelihood enters as an effective observation field, enabling joint inference of latent activity, network couplings, and emission parameters. The framework extends naturally to maximum-a-posteriori inference with structured priors, including sparsity and a hierarchical extension favouring Dale-consistent outgoing interactions.

We validate parameter recovery on matched synthetic data and test the method on spike trains generated by a recurrent conductance-based leaky integrate-and-fire network. In this setting, SpiKIsing recovers sparse connection structure and correctly classifies all excitatory and inhibitory neurons despite
the substantial mismatch between the data-generating dynamics and the SpiKIsing model.
\end{abstract}

\section{Introduction}

Inferring the interactions underlying the collective dynamics of neuronal populations from recorded spike trains is a fundamental inverse problem in computational neuroscience and non-equilibrium statistical physics \cite{stevenson2008inferring,nguyen2017inverse}. Modern electrophysiological techniques allow for the simultaneous recording of increasingly large neuronal populations with high temporal resolution \cite{jun2017fully,steinmetz2021neuropixels}, raising the question of how effective interactions between neurons can be reconstructed from their observed spiking activity.

This inference problem is intrinsically difficult. The interactions are not observed directly, neuronal activity is stochastic and strongly history dependent, and spike trains provide only an indirect manifestation of the underlying network dynamics. Correlation- and covariance-based approaches can reveal statistical dependencies between spike trains, but such dependencies may arise from direct interactions, common inputs, or indirect network effects and do not by themselves define a generative dynamical model \cite{stevenson2008inferring,friston2011functional,ostojic2009connectivity}. Point-process and generalized-linear-model approaches instead describe spike generation directly in continuous time and can naturally incorporate coupling filters, stimulus dependence, and single-neuron spike-history effects \cite{truccolo2005point,pillow2008spatio}.

A complementary class of approaches, originating from statistical physics, represents neuronal activity through binary variables and infers their interactions using Ising-type models \cite{schneidman2006weak,cocco2009neuronal,roudi2009ising}. Their dynamical extension, the kinetic Ising model, is particularly suited to directed non-equilibrium interactions and has been widely used to reconstruct effective connectivity from time-dependent neuronal activity \cite{roudi2011mean,mezard2011exact,zeng2011network,donner2017inverse,po2025inferring}. When applied to spike trains, however, kinetic Ising inference typically requires discretizing continuous spike times into binary time bins \cite{hertz2011ising,tyrcha2013effect,capone2015inferring,po2025inferring}. The resulting representation introduces an analysis timescale and directly identifies observed spike events with the dynamical variables of the Ising model. Large bins can merge distinct temporal events, whereas small bins lead to increasingly sparse activity and still discard within-bin spike timing. More importantly, this identification mixes collective network dynamics with single-neuron temporal effects, such as refractoriness and post-spike recovery, which may then be spuriously attributed to network interactions.

Here we introduce the \textbf{Spiking-Kinetic-Ising} (\textit{SpiKIsing}) model, a latent-variable framework that separates these two distinct but interacting levels. An \textit{unobserved} binary network state $\vect{s}(t)$ evolves according to an asymmetric kinetic Ising model, while the \textit{observed} spike trains are generated in continuous time by history-dependent point processes conditioned on the latent trajectory. The interaction matrix $J$ therefore describes the directed collective dynamics between neurons, whereas the emission process describes how each latent state is expressed as individual spikes and accounts for single-neuron spike-history effects. In the specific model considered here, the emission process is a history-dependent point process whose intensity distinguishes the active and inactive latent states and incorporates an absolute refractory period together with a post-spike recovery function.

Latent state-space models with point-process observations have previously been used to infer unobserved dynamical processes from neural spike trains~\cite{smith2003estimating,kulkarni2007common}. SpiKIsing differs from these approaches in that its latent variables form an interacting binary network, whose directed couplings are themselves objects of inference. Related work has also considered partially observed kinetic Ising models, in which a subset of the network spins is hidden~\cite{dunn2013learning,battistin2015belief}. Here, by contrast, a latent dynamical state is associated with every recorded neuron and mediates between the interacting network dynamics and the continuous-time spike observations.

The latent representation makes exact inference intractable, since the likelihood requires summing over all possible dynamical trajectories. We therefore derive a variational mean-field approximation and an expectation-maximization scheme for jointly inferring the latent activity, interaction matrix, local fields, and emission parameters. A central result is that the continuous-time point-process likelihood enters the mean-field equations through an effective \textit{observation field}, given by the log-likelihood ratio between the active and inactive latent states. This preserves the structure of kinetic-Ising inference while allowing the precise spike times to contribute directly to the likelihood. We derive both a forward--backward variational update and a computationally cheaper forward-only approximation.

Biological networks also exhibit structural organization that is not explicitly enforced by the unconstrained model. The same framework also allows prior structural information about the interaction network to be incorporated through maximum-a-posteriori inference. Specifically, we consider a sparsity-promoting prior and a hierarchical extension motivated by Dale's principle~\cite{strata1999dale}, in which the excitatory or inhibitory identity of each presynaptic neuron is inferred jointly with its outgoing interactions and controls their preferred sign. We first validate parameter recovery in matched synthetic data and then assess the method on spike trains generated by recurrent conductance-based
leaky integrate-and-fire network, for which the recurrent connectivity and
presynaptic excitatory or inhibitory identities are known, while the
spike-generating dynamics lies outside the SpiKIsing model class.

The paper is organized as follows. In Sec.~\ref{sec:Model} we introduce the SpiKIsing generative model, and in Sec.~\ref{sec:Inference} we derive the variational inference scheme. Section~\ref{sec:MAP} introduces structured priors for sparse and Dale-consistent connectivity, while Sec.~\ref{sec:LIF} tests the method on leaky integrate-and-fire networks. Finally, Sec.~\ref{sec:Conclusions} discusses the limitations of the present approach, its scope, and possible extensions.

\section{\label{sec:Model}The SpiKIsing Model}
The SpiKIsing model separates the dynamics into two layers: a discrete-time latent process describing collective network activity and a continuous-time emission process describing spike generation conditioned on the latent state.

We consider a system of $N$ neurons with latent state
$$
\vect{s}(t)=\bigl(s_1(t),\ldots,s_N(t)\bigr),
\qquad
s_i(t)\in\{0,1\},
$$
at discrete times $t\in\{0,\ldots,T-1\}$, where $s_i(t)=1$ and $s_i(t)=0$ denote active and inactive latent states, respectively. We write the complete latent trajectory as
$$
\vect{s}_{0:T-1}=\{\vect{s}(t)\}_{t=0}^{T-1}.
$$

Continuous time is measured in units of one latent time step, so that $\vect{s}(t)$ is associated with the interval $[t,t+1)$.
For each neuron $i$, the observations consist of the continuous-time spike train
$$
\vect{\tau}_i=\{\tau_{i,k}\}_{k=1}^{n_i},
\qquad
0<\tau_{i,1}<\cdots<\tau_{i,n_i}<T,
$$
and we denote the complete set of recordings by
$\vect{\tau}=\{\vect{\tau}_i\}_{i=1}^N$.

We separate the parameters into
$\theta_{\mathrm{lat}}$, governing the latent dynamics, and
$\theta_{\mathrm{obs}}$, regulating the emission process and hence spike generation. The marginal likelihood is then
\begin{equation}
    P(\vect{\tau}\mid\theta_{\mathrm{lat}},\theta_{\mathrm{obs}})
    =
    \sum_{\vect{s}_{0:T-1}}
    P(\vect{\tau}\mid\vect{s}_{0:T-1},\theta_{\mathrm{obs}})
    P(\vect{s}_{0:T-1}\mid\theta_{\mathrm{lat}}).
    \label{eq:Likel_tau}
\end{equation}

Conditional on the latent trajectory, spike generation is independent across neurons,
\begin{equation}
P(\vect{\tau}\mid\vect{s}_{0:T-1},\theta_{\mathrm{obs}})
=
\prod_{i=1}^N
P(\vect{\tau}_i\mid\vect{s}_i,\theta_{\mathrm{obs}}),
\end{equation}
where $\vect{s}_i=\{s_i(t)\}_{t=0}^{T-1}$. Correlations between observed spike trains therefore arise through the interacting latent dynamics.

\subsection{Latent dynamics}
We model the latent network by an \textbf{asymmetric kinetic Ising model}~\cite{roudi2011dynamical,mezard2011exact} with first-order Markov dynamics,
\begin{equation}
P(\vect{s}_{0:T-1}\mid\theta_{\mathrm{lat}})
=
P(\vect{s}(0))
\prod_{t=1}^{T-1}
P\!\left(\vect{s}(t)\mid\vect{s}(t-1)\right),
\end{equation}
and conditionally independent updates,
\begin{equation}
P\!\left(\vect{s}(t)\mid\vect{s}(t-1)\right)
=
\prod_{i=1}^N
P\!\left(s_i(t)\mid\vect{s}(t-1)\right).
\end{equation}

For the $0/1$ representation used here,
\begin{equation}
P\!\left(s_i(t)\mid\vect{s}(t-1)\right)
=
\frac{\exp\!\left[s_i(t)H_i(t)\right]}
     {1+\exp H_i(t)},
\end{equation}
with local field
\begin{equation}
H_i(t)=h_i+\sum_{j=1}^N J_{ij}s_j(t-1).
\end{equation}

Equivalently,
\begin{equation}
P\!\left(s_i(t)=1\mid\vect{s}(t-1)\right)
=
\sigma\!\left(H_i(t)\right),
\end{equation}
where $\sigma(x)=(1+e^{-x})^{-1}$ is the sigmoid function.

The field $h_i$ sets the baseline tendency of neuron $i$ to occupy the active state, while $J_{ij}$ describes the directed influence of neuron $j$ at the preceding time step. Positive and negative couplings respectively increase and decrease the probability of activation of neuron $i$.

We impose $ J_{ii}=0 $, so that $J$ contains only interactions between distinct neurons, while spike-history effects are represented explicitly in the emission process. The couplings are otherwise unrestricted and may be asymmetric, $J_{ij}\neq J_{ji}$, as expected for directed interactions. Such asymmetry generically breaks detailed balance, placing the latent dynamics in the class of non-equilibrium stochastic systems \cite{crisanti1988dynamics,aguilera2023nonequilibrium}. We assume throughout that $J$ and $\vect h$ are time independent, such that $\theta_{\mathrm{lat}}=\{J,\vect h\}$ contains $N^2$ parameters in total.

\subsection{Spike-train emission model}

The emission layer maps each latent trajectory to a continuous-time spike train. We model this conditional mapping as a history-dependent point process whose intensity is modulated by the current latent state.
For continuous time $\tau$, we define
$$
s_i(\tau)\equiv s_i(\lfloor\tau\rfloor),
$$
so that the latent state is constant within each interval $[t,t+1)$. Let
\begin{equation}
    \tau_i^*(\tau)
    =
    \max\{\tau_{i,k}:\tau_{i,k}<\tau\} ,
\end{equation}
denoting the most recent spike preceding time $\tau$. Following at least one observed spike, we define the elapsed time beyond the absolute refractory period as
\begin{equation}
    \Delta_i(\tau)
    =
    \tau-\tau_i^*(\tau)-\tau_i^{\rm abs}.
\end{equation}

Before the first observed spike we set
$\Delta_i(\tau)=+\infty$, corresponding to a fully recovered initial condition.
Given the current latent state and the preceding spike history, the
point-process intensity is
\begin{equation}
\lambda_i(\tau)
=
\begin{cases}
0,
&
\Delta_i(\tau)\leq0,
\\[4pt]
\lambda_i^-,
&
\Delta_i(\tau)>0,\quad s_i(\tau)=0,
\\[4pt]
\lambda_i^+\rho_i\!\left(\Delta_i(\tau)\right),
&
\Delta_i(\tau)>0,\quad s_i(\tau)=1.
\end{cases}
\end{equation}

Here $\lambda_i^+$ and $\lambda_i^-$ are the firing rates associated with the active and inactive latent states, respectively. We generally consider
$\lambda_i^-\ll\lambda_i^+$, while retaining a non-zero inactive-state rate. Following the absolute refractory period, neuronal excitability typically recovers gradually \cite{berry1997refractoriness}. We represent this effect using the phenomenological one-parameter kernel
\begin{equation}
    \rho_i(\delta)
    =
    \frac{\delta}{\delta+c_i},
    \qquad
    \delta>0,
    \label{eq:recovery}
\end{equation}
chosen for its monotonic recovery and analytical tractability. Here $c_i\geq0$ controls the recovery timescale. For $c_i>0$,
$\rho_i(c_i)=1/2$, while $\rho_i(\delta)\to1$ for $\delta\gg c_i$.
The limiting case $c_i=0$ gives $\rho_i(\delta)=1$ for all $\delta>0$,
corresponding to instantaneous recovery.

For a simple point process~\cite{truccolo2005point}, the conditional log-likelihood is

\begin{equation}
\label{eq:LikelihoodPP}
\ln P(\vect{\tau}_i\mid\vect{s}_i,\theta_{\mathrm{obs}})
=
\sum_{k=1}^{n_i}\ln\lambda_i(\tau_{i,k})
-
\int_0^T\lambda_i(u)\,du.
\end{equation}

The latent state therefore modulates the intensity over each discrete interval, while the likelihood retains the continuous observed spike times.

The emission parameters are

$$
\theta_{\mathrm{obs}}
=
\{
\vect{\lambda}^+,
\vect{\lambda}^-,
\vect c,
\vect{\tau}^{\rm abs}
\}.
$$

\begin{figure*}
    \centering
    \includegraphics[width=\textwidth]{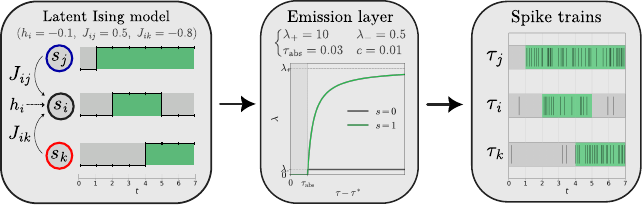}
    \caption{
    \textbf{Schematic representation of the SpiKIsing model.}
    \textbf{(Left)} Latent trajectories for three neurons. Neuron $i$ receives
    positive coupling $J_{ij}=0.5$ from neuron $j$, negative coupling
    $J_{ik}=-0.8$ from neuron $k$, and local field $h_i=-0.1$.
    The state of the presynaptic neurons at time $t-1$ determines the probability that neuron $i$ occupies its active latent state at time $t$.
    \textbf{(Centre)} Emission layer, represented by the conditional point-process intensity of neuron $i$, with
    $\lambda_i^+=10$, $\lambda_i^-=0.5$,
    $\tau_i^{\rm abs}=0.03$, and $c_i=0.01$.
    Following the absolute refractory interval, the active-state intensity
    recovers according to $\rho_i(\delta)=\delta/(\delta+c_i)$.
    \textbf{(Right)} A stochastic realization of the continuous-time spike
    trains generated from the latent trajectories (left), through the emission layer (centre).
    }
    \label{fig:Scheme}
\end{figure*}

Figure~\ref{fig:Scheme} illustrates the complete generative mechanism. Positive and negative incoming couplings respectively favour and suppress the activation of neuron $i$ at the following latent time step. Conditional on the resulting latent trajectory, the emission layer generates individual spikes stochastically: active intervals are associated with a higher firing rate, while refractoriness and post-spike recovery determine the short-time structure following each spike. Thus the latent trajectory controls the collective dynamical state, but does not uniquely determine the observed spike times.

\subsection{Likelihood for a single time interval}

For the variational inference developed below, it is convenient to decompose
the observation likelihood into contributions from individual latent intervals.
Let
$$
\vect{\tau}_i^t
=
\{\tau_{i,k}:t\leq\tau_{i,k}<t+1\},
\qquad
n_i^t=|\vect{\tau}_i^t|,
$$
denote the spikes of neuron $i$ in $[t,t+1)$. Because the emission process is
history dependent, the likelihood in a given interval also depends on the
observed spike history preceding that interval. We denote this history by
\[
\mathcal H_i(t)
=
\{\tau_{i,k}:\tau_{i,k}<t\}.
\]
For compactness, below we suppress the conditioning on $\mathcal H_i(t)$ and
on the fixed emission parameters, writing
$P(\vect{\tau}_i^t\mid s_i(t))$ as shorthand for
$P(\vect{\tau}_i^t\mid s_i(t),\mathcal H_i(t),\theta_{\rm obs})$.
Since $s_i(\tau)=s_i(t)$ throughout the interval,
\begin{equation}
\ln P(\vect{\tau}_i\mid\vect{s}_i)
=
\sum_{t=0}^{T-1}
\ln P(\vect{\tau}_i^t\mid s_i(t)).
\end{equation}

We define the non-refractory exposure
\begin{equation}
E_i^t
=
\int_t^{t+1}
\mathbb I\!\left[\Delta_i(u)>0\right]\,du,
\end{equation}
the recovery-weighted exposure
\begin{equation}
A_i^t
=
\int_t^{t+1}
\rho_i\!\left(\Delta_i(u)\right)
\mathbb I\!\left[\Delta_i(u)>0\right]\,du,
\end{equation}
and the recovery contribution from observed spikes,
\begin{equation}
    C_i^t
    =
    \sum_{\tau_{i,k}\in[t,t+1)}
    \log \rho_i\!\left(\Delta_i(\tau_{i,k})\right).
\end{equation}
Because $\Delta_i$ is defined from the most recent spike in the complete
recording, spike history is carried continuously across latent-interval
boundaries. Under the fully recovered initial condition,
$\Delta_i(\tau_{i,1})=+\infty$, so the first recorded spike contributes
$\log\rho_i(+\infty)=0$.
For the recovery kernel in
Eq.~\eqref{eq:recovery}, $E_i^t$, $A_i^t$, and $C_i^t$ can all be
evaluated analytically from the observed spike times; the explicit expressions,
together with the derivatives required for inference, are given in
Appendix~\ref{app:DerC}.

The interval log-likelihood is therefore
\begin{equation}
\ln P(\vect{\tau}_i^t\mid s_i(t))
=
\begin{cases}
n_i^t\ln\lambda_i^-
-
\lambda_i^-E_i^t,
&
s_i(t)=0,
\\[6pt]
n_i^t\ln\lambda_i^+
+
C_i^t
-
\lambda_i^+A_i^t,
&
s_i(t)=1.
\end{cases}
\end{equation}

These quantities depend only on the observed spike history and the emission parameters. In Sec.~\ref{sec:Inference}, their difference between the two
latent states will define the effective observation field entering the
mean-field equations.

\section{Variational inference}
\label{sec:Inference}

Having defined the generative model, we now consider the inverse problem of
estimating its parameters from the observed spike trains. We first focus on
maximum-likelihood inference and seek
\begin{equation}
    \{\theta_{\rm obs}^*,\theta_{\rm lat}^*\}
    =
    \operatorname*{argmax}_{\theta_{\rm obs},\theta_{\rm lat}}
    P(\vect{\tau}\mid\theta_{\rm obs},\theta_{\rm lat}).
    \label{eq:ML_objective}
\end{equation}
The marginal likelihood in Eq.~\eqref{eq:Likel_tau} requires a sum over
$2^{NT}$ possible latent trajectories and is therefore intractable already
for moderately sized systems. We address this by introducing a variational
mean-field approximation to the posterior distribution over the latent states.

\subsection{Mean-field approximation and variational free energy}

For fixed model parameters, we approximate the posterior over latent
trajectories by a fully factorized distribution
\begin{equation}
    Q(\vect{s}_{0:T-1})
    =
    \prod_{i=1}^N\prod_{t=0}^{T-1}Q_i^t(s_i(t)),
    \qquad
    Q_i^t(s_i(t)=1)=m_i(t),
    \label{eq:Q_MF}
\end{equation}
with $Q_i^t(s_i(t)=0)=1-m_i(t)$. The variational parameters
$m_i(t)\in[0,1]$ approximate the posterior probability that neuron $i$
occupies its active latent state at time $t$.

The standard variational construction \cite{jordan1999introduction,blei2017variational}, recalled in
Appendix~\ref{app:Var}, gives the lower bound
\begin{equation}
    \log P(\vect{\tau}\mid\theta_{\rm obs},\theta_{\rm lat})
    \geq
    \mathcal F[Q]
    =
    \mathbb E_Q
    \left[
        \log P(\vect{\tau},\vect{s}_{0:T-1}
        \mid\theta_{\rm obs},\theta_{\rm lat})
    \right]
    +
    \mathcal S_Q,
    \label{eq:elbo_def}
\end{equation}
where $\mathcal S_Q=-\mathbb E_Q[\log Q]$ is the entropy of the
variational distribution. Maximizing $\mathcal F$ therefore provides a
tractable approximation to maximum-likelihood inference; the bound becomes
tight when $Q$ coincides with the exact posterior.

Taking the initial latent state to be uniform, its contribution is constant
and the variational objective separates as
\begin{equation}
    \mathcal F
    =
    \mathcal L_{\rm obs}
    +
    \mathcal L_{\rm lat}
    +
    \mathcal S_Q,
    \label{eq:elbo_components}
\end{equation}
with
\begin{align}
    \mathcal L_{\rm obs}
    &=
    \sum_{i,t}
    \mathbb E_{Q_i^t}
    \left[
        \log P(\vect{\tau}_i^t\mid s_i(t))
    \right],
    \label{eq:Lobs_def}
    \\
    \mathcal L_{\rm lat}
    &=
    \sum_i\sum_{t=1}^{T-1}
    \mathbb E_Q
    \left[
        s_i(t)H_i(t)
        -
        \log\left(1+\exp H_i(t)\right)
    \right].
    \label{eq:Llatent_def}
\end{align}
Using the interval likelihood derived in Sec.~\ref{sec:Model}, the observation
term is evaluated exactly under the factorized distribution,
\begin{align}
    \mathcal L_{\rm obs}
    =
    \sum_{i,t}
    \Big[
        &(1-m_i(t))
        \left(
            n_i^t\log\lambda_i^-
            -
            \lambda_i^-E_i^t
        \right)
        \nonumber\\
        &+
        m_i(t)
        \left(
            n_i^t\log\lambda_i^+
            +
            C_i^t
            -
            \lambda_i^+A_i^t
        \right)
    \Big],
    \label{eq:Lobs_explicit}
\end{align}
while
\begin{equation}
    \mathcal S_Q
    =
    -\sum_{i,t}
    \left[
        m_i(t)\log m_i(t)
        +
        (1-m_i(t))\log(1-m_i(t))
    \right].
    \label{eq:entropy_MF}
\end{equation}

The latent contribution remains non-trivial because the local fields fluctuate
under $Q$. Introducing their mean
\begin{equation}
    \eta_i(t)
    =
    h_i+\sum_jJ_{ij}m_j(t-1),
    \label{eq:eta_def}
\end{equation}
the linear contribution satisfies
$\mathbb E_Q[s_i(t)H_i(t)]=m_i(t)\eta_i(t)$ exactly. At the naive
mean-field (nMF) level, we approximate the remaining nonlinear term by
\begin{equation}
    \mathbb E_Q
    \left[
        \log(1+\exp H_i(t))
    \right]
    \simeq
    \log(1+\exp\eta_i(t)).
\end{equation}
This gives
\begin{equation}
    \mathcal L_{\rm lat}^{\rm nMF}
    =
    \sum_i\sum_{t=1}^{T-1}
    \left[
        m_i(t)\eta_i(t)
        -
        \log(1+\exp\eta_i(t))
    \right],
    \label{eq:elbo_latent_nmf}
\end{equation}
and we henceforth optimize
\begin{equation}
    \mathcal F_{\rm nMF}
    =
    \mathcal L_{\rm obs}
    +
    \mathcal L_{\rm lat}^{\rm nMF}
    +
    \mathcal S_Q.
    \label{eq:F_nMF}
\end{equation}

It is important to distinguish the two approximations involved here.
Equation~\eqref{eq:elbo_def} is a rigorous variational lower bound, whereas
$\mathcal F_{\rm nMF}$ includes the additional replacement of the fluctuating
local field by its mean and is therefore not guaranteed to remain a strict
lower bound. Higher-order mean-field corrections could be incorporated at
this stage \cite{roudi2011dynamical,mezard2011exact,bachschmid2016variational}; here we use the nMF approximation to retain a computationally
simple inference scheme.

\subsection{Variational expectation-maximization algorithm}

We optimize $\mathcal F_{\rm nMF}$ using a variational
expectation-maximization scheme \cite{dempster1977maximum,neal1998view}. The E-step updates the variational
parameters $m_i(t)$ for fixed model parameters, while the M-step updates the
model parameters for fixed $Q$. The full E-step seeks a stationary point of
$\mathcal F_{\rm nMF}$ with respect to the magnetizations; below we also
introduce a computationally cheaper forward-only update, which constitutes an
additional approximation.

\subsubsection{E-step}

The dependence of the observation likelihood on $m_i(t)$ can be summarized by
an effective observation field,
\begin{align}
    h_i^{\rm obs}(t)
    &\equiv
    \frac{\partial\mathcal L_{\rm obs}}{\partial m_i(t)}
    =
    \log
    \frac{
        P(\vect{\tau}_i^t\mid s_i(t)=1)
    }{
        P(\vect{\tau}_i^t\mid s_i(t)=0)
    }
    \nonumber\\
    &=
    n_i^t\log\frac{\lambda_i^+}{\lambda_i^-}
    +
    C_i^t
    -
    \lambda_i^+A_i^t
    +
    \lambda_i^-E_i^t.
    \label{eq:h_obs_continuous}
\end{align}
Thus $h_i^{\rm obs}(t)$ is the log-likelihood ratio contributed by the
continuous-time spike train during the interval $[t,t+1)$: positive values
favour the active latent state and negative values the inactive state. In this
form, the complete information supplied by the point-process emission model
to the latent Ising variable at time $t$ is represented by a single effective
field.

Differentiating $\mathcal F_{\rm nMF}$ with respect to $m_i(t)$ and imposing
stationarity gives, for $1\leq t\leq T-2$,
\begin{equation}
    m_i(t)
    =
    \sigma
    \left[
        \eta_i(t)
        +
        h_i^{\rm obs}(t)
        +
        R_i(t)
    \right],
    \label{eq:mean_field_exact}
\end{equation}
where
\begin{equation}
    R_i(t)
    =
    \sum_kJ_{ki}
    \left[
        m_k(t+1)
        -
        \sigma\left(\eta_k(t+1)\right)
    \right].
    \label{eq:reaction}
\end{equation}
The reaction term $R_i(t)$ arises because $m_i(t)$ contributes to the local
fields at the following time step and therefore propagates information
backwards from future latent states. At the final time step
$R_i(T-1)=0$, while at $t=0$ the uniform initial-state prior removes the
contribution from the preceding latent dynamics.

Because Eq.~\eqref{eq:mean_field_exact} couples neighbouring times in both
directions, we solve it iteratively using forward--backward sweeps. The
forward pass evaluates the mean local fields $\eta_i(t)$, while the backward
pass propagates $R_i(t)$ and updates the magnetizations; the procedure is
repeated until convergence.

For greater computational efficiency, we also consider a causal
forward-filtering approximation in which the backward reaction term is
neglected,
\begin{equation}
    m_i(t)
    =
    \sigma
    \left[
        \eta_i(t)
        +
        h_i^{\rm obs}(t)
    \right],
    \qquad t\geq1,
    \label{eq:mean_field_forward}
\end{equation}
with $m_i(0)=\sigma(h_i^{\rm obs}(0))$. This update requires only a single
forward pass, but unlike Eq.~\eqref{eq:mean_field_exact} it is not a
stationary solution of $\mathcal F_{\rm nMF}$. In
Appendix~\ref{app:FBvsB} we compare the two schemes empirically and quantify
the accuracy lost by neglecting the backward term.

\subsubsection{M-step}

In the M-step, the magnetizations $m_i(t)$ are held fixed and
$\mathcal F_{\rm nMF}$ is optimized with respect to the model parameters.

For the latent parameters $\theta_{\rm lat}=\{J,\vect h\}$, the coupling
gradient is
\begin{equation}
    \frac{\partial\mathcal F_{\rm nMF}}{\partial J_{ij}}
    =
    \sum_{t=1}^{T-1}
    m_j(t-1)
    \left[
        m_i(t)
        -
        \sigma(\eta_i(t))
    \right].
    \label{eq:grad_J}
\end{equation}
The local fields $h_i$ are updated by directly imposing their stationarity
condition,
\begin{equation}
    \sum_{t=1}^{T-1}
    \sigma
    \left(
        h_i+\sum_jJ_{ij}m_j(t-1)
    \right)
    =
    \sum_{t=1}^{T-1}m_i(t),
    \label{eq:h_root_finding}
\end{equation}
which can be solved independently for each neuron by one-dimensional
root finding.

In practice, the $0/1$ parametrization can lead to poor conditioning because
the presynaptic magnetizations have non-zero temporal means. In all numerical
experiments we therefore use the algebraically equivalent centred
parametrization described in Appendix~\ref{app:centered-reparam}, which
leaves the mean fields $\eta_i(t)$ and hence the model likelihood unchanged.

For the emission parameters, the active- and inactive-state firing rates
admit closed-form updates,
\begin{align}
    \lambda_i^+
    &=
    \frac{
        \sum_t m_i(t)n_i^t
    }{
        \sum_t m_i(t)A_i^t
    },
    \label{eq:update_lam_plus}
    \\
    \lambda_i^-
    &=
    \frac{
        \sum_t(1-m_i(t))n_i^t
    }{
        \sum_t(1-m_i(t))E_i^t
    }.
    \label{eq:update_lam_minus}
\end{align}
The recovery parameter $c_i$ is updated by gradient ascent. For the kernel in
Eq.~\eqref{eq:recovery}, the required derivatives of $A_i^t$ and $C_i^t$
are available analytically and are given in Appendix~\ref{app:DerC}.

The absolute refractory time $\tau_i^{\rm abs}$ may either be fixed when
independent information is available or estimated by one-dimensional profile
maximization of the observation likelihood. The corresponding procedure is
described in Appendix~\ref{app:DerC}. In the matched-model experiments of
Secs.~\ref{sec:matched_validation} and \ref{sec:structured_validation}, we
fix $\tau_i^{\rm abs}$ to its generating value; it is instead inferred in the
leaky integrate-and-fire experiments of Sec.~\ref{sec:LIF}.

\subsection{Matched-model validation}
\label{sec:matched_validation}

Before introducing structured priors or testing the method under more realistic
spike-generating dynamics, we first consider a controlled setting in which
the data are generated directly from the SpiKIsing model. This allows us to
assess parameter recovery while isolating finite-data effects and the
approximations introduced by the inference procedure.

Unless stated otherwise, the numerical results in this subsection use the
forward-only variational update of Eq.~\eqref{eq:mean_field_forward} and the
centred M-step parametrization of Appendix~\ref{app:centered-reparam}. We use networks of $N=100$ neurons with off-diagonal
couplings
\begin{equation}
    J_{ij}
    \sim
    \mathcal N\left(0,\frac{g^2}{N}\right),
    \qquad i\neq j,
\end{equation}
where $g$ controls the overall coupling scale; we set $g=1$ unless varied
explicitly. The local fields are drawn independently
from a Gaussian distribution with mean $-0.4$ and variance $0.02$, while the
emission parameters are
$\lambda_i^+=10$, $\lambda_i^-=0.01$, $c_i=0.2$, and
$\tau_i^{\rm abs}=0.01$ for all neurons. In these matched-model experiments,
$\tau_i^{\rm abs}$ is fixed to its generating value, while the remaining
parameters are inferred unless stated otherwise. The default recording length
is $T=50000$.

\begin{figure}
    \centering
    \begin{subfigure}[t]{0.48\linewidth}
        \centering
        \includegraphics[width=\linewidth]{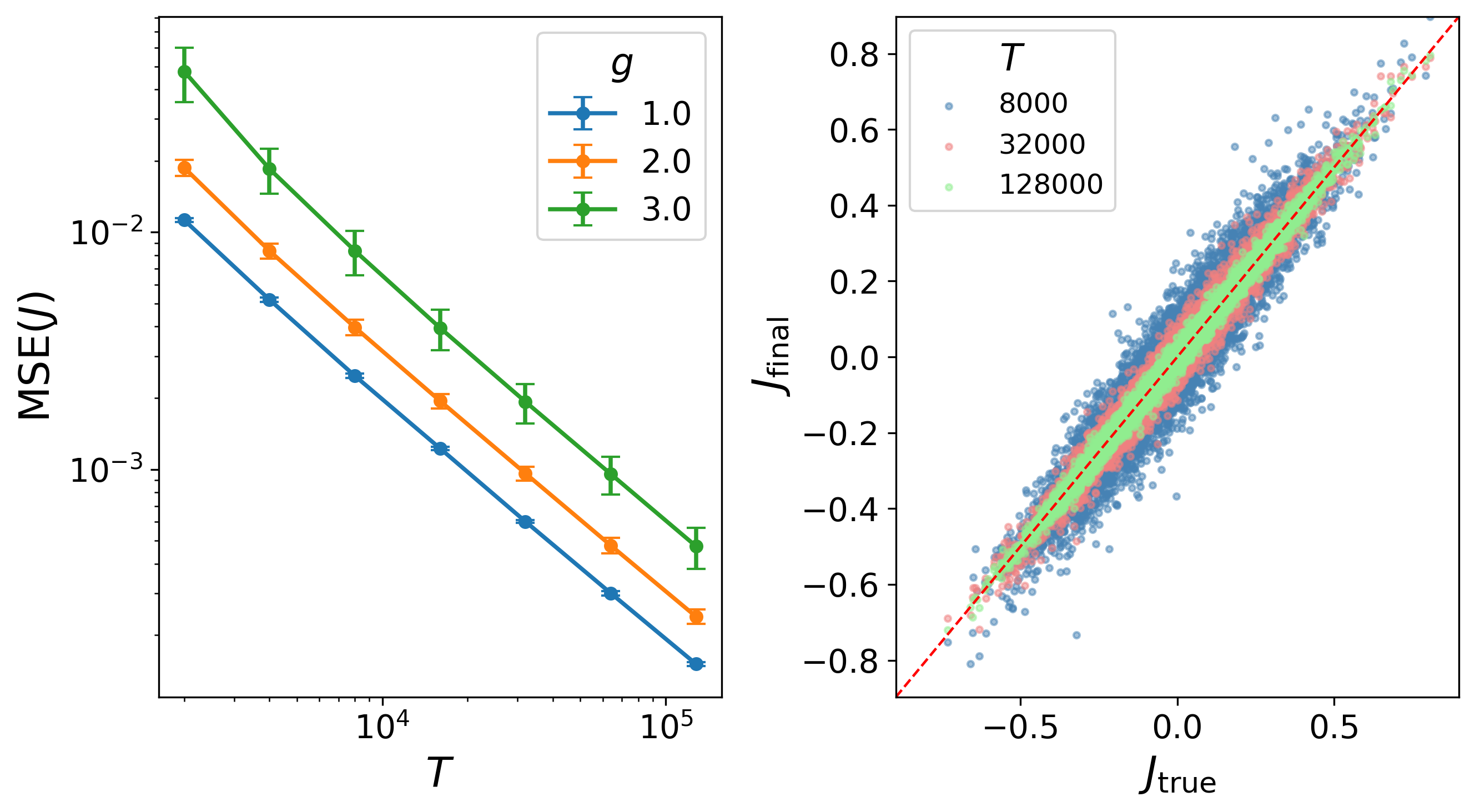}
        \caption{}
        \label{fig:MSEJ}
    \end{subfigure}\hfill
    \begin{subfigure}[t]{0.48\linewidth}
        \centering
        \includegraphics[width=\linewidth]{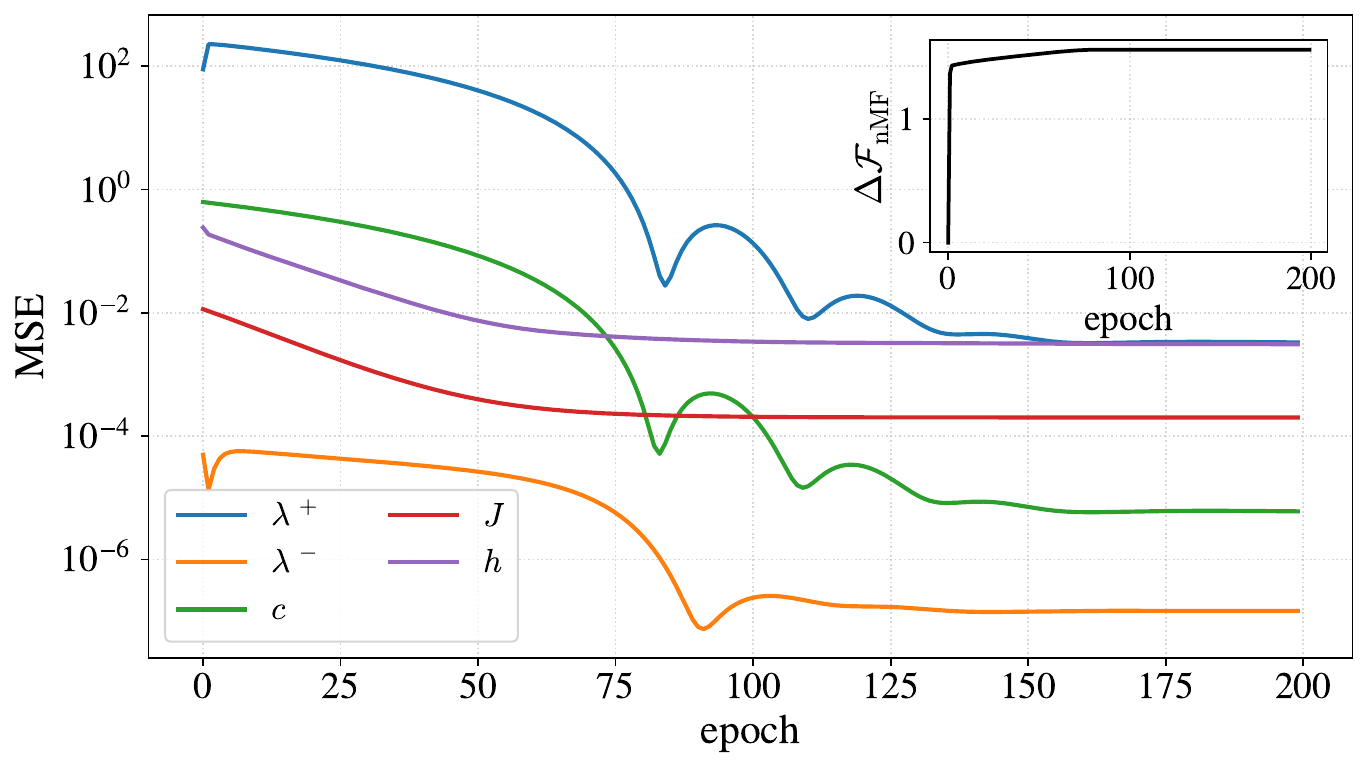}
        \caption{}
        \label{fig:Learning}
    \end{subfigure}
    \caption{
    \textbf{Matched-model validation of maximum-likelihood SpiKIsing inference.} \\
    \textbf{(a)} \textbf{Interaction recovery.}
        Left: mean-squared error of the inferred interaction matrix as a
        function of the recording length $T$ for different coupling strengths
        $g$. Right: inferred versus ground-truth couplings for three
        representative recording lengths. \textbf{(b)} \textbf{Joint learning of latent and emission parameters.}
        Evolution of the mean-squared errors in $J$, $\vect h$, $\vect\lambda^+$,
        $\vect\lambda^-$, and $\vect c$ during inference. The inset shows
        $\Delta\mathcal F_{\rm nMF}
        =
        \mathcal F_{\rm nMF}({\rm epoch})-\mathcal F_{\rm nMF}(0)$.
    Simulation parameters are specified in the text.
    }
    \label{fig:GaussianLearning}
\end{figure}

We quantify interaction recovery using
\begin{equation}
    {\rm MSE}(J)
    =
    \frac{1}{N(N-1)}
    \sum_{i\neq j}
    \left(
        J_{ij}^{\rm inf}-J_{ij}^{\rm true}
    \right)^2.
    \label{eq:MSE_J}
\end{equation}
For $\vect h$, $\vect\lambda^+$, $\vect\lambda^-$, and $\vect c$, the
mean-squared error is defined analogously by averaging the squared
component-wise difference between inferred and ground-truth values.
Figure~\ref{fig:MSEJ} shows that the reconstruction error decreases
systematically with recording length for all coupling strengths considered.
Consistently, the inferred couplings concentrate progressively around
$J_{ij}^{\rm inf}=J_{ij}^{\rm true}$ as more data become available.

We next test joint learning of the latent parameters and of the emission
parameters $\vect\lambda^+$, $\vect\lambda^-$, and $\vect c$, initializing
the algorithm away from their ground-truth values. Figure~\ref{fig:Learning}
shows that the errors in $J$, $\vect h$, $\vect\lambda^+$,
$\vect\lambda^-$, and $\vect c$ decrease and eventually stabilize, although
the different parameter families converge on different timescales and can
exhibit transient non-monotonic behaviour. The absolute refractory times are
kept fixed at their generating values in this experiment. The simultaneous
increase and saturation of $\mathcal F_{\rm nMF}$ provides an additional
diagnostic of convergence.

\section{\label{sec:MAP}MAP inference with structured priors}

The inference framework developed in Sec.~\ref{sec:Inference} estimates the
model parameters by maximum likelihood. Structural information about the
effective interaction network can instead be incorporated through
maximum-a-posteriori (MAP) inference \cite{donner2017inverse,po2025inferring}. We introduce a prior $P(J)$ on the
coupling matrix, while retaining the maximum-likelihood treatment of
$\vect h$ and the emission parameters, so that
\begin{equation}
    P(J,\vect h,\theta_{\rm obs}\mid\vect{\tau})
    \propto
    P(\vect{\tau}\mid J,\vect h,\theta_{\rm obs})P(J).
    \label{eq:posterior_J}
\end{equation}
The MAP estimate maximizes this posterior, balancing agreement with the
observed spike trains against the structural information encoded by $P(J)$.
At the nMF level, the corresponding objective is
\begin{equation}
    \mathcal F_{\rm nMF}^{\rm MAP}
    =
    \mathcal F_{\rm nMF}
    +
    \log P(J).
    \label{eq:F_nMF_MAP}
\end{equation}
Because the prior acts only on the interaction matrix, the variational E-step
for the dynamical latent states is unchanged, while the coupling update in
the M-step is modified according to the chosen prior. Below we first introduce a Laplace prior promoting sparse effective connectivity, and then extend it hierarchically to favour sign consistency
among the outgoing interactions of each presynaptic neuron, as motivated by
Dale's principle~\cite{strata1999dale}.

\subsection{Sparse connectivity}
\label{sec:sparsity}

Many neuronal circuits are anatomically sparse \cite{song2005highly,lefort2009excitatory}, motivating a sparse
representation also for the inferred effective interaction network. Sparsity-promoting regularization has previously been applied to connectivity
reconstruction in non-equilibrium kinetic Ising models, particularly through
$L_1$ penalties~\cite{zeng2011network}.
We impose a Laplace prior on the off-diagonal couplings,
\begin{equation}
    P(J)
    \propto
    \exp\left[
        -\gamma_1
        \sum_{i\neq j}|J_{ij}|
    \right],
    \label{eq:l1_prior}
\end{equation}
which adds an $L_1$ penalty to the nMF objective and thereby implements
the same sparsity-promoting mechanism underlying Lasso estimation~\cite{tibshirani1996regression}.

Because the $L_1$ term is non-differentiable at zero, we optimize the
resulting non-smooth objective using a proximal-gradient update~\cite{parikh2014proximal}. Let
\begin{equation}
    \widetilde J_{ij}
    =
    J_{ij}
    +
    \alpha G_{ij}
\end{equation}
denote a gradient-ascent step on the smooth likelihood contribution, where $\alpha>0$ is the learning rate and $G_{ij}$ denotes the
centred coupling gradient defined in
Appendix~\ref{app:centered-reparam}. The coupling is then updated through
soft thresholding,
\begin{equation}
    J_{ij}^{\rm new}
    =
    \operatorname{sign}(\widetilde J_{ij})
    \max\left(
        0,
        |\widetilde J_{ij}|-\alpha\gamma_1
    \right).
    \label{eq:proximal_operator}
\end{equation}
The $L_1$ prior can therefore set weak inferred interactions exactly to zero,
without requiring a separate post-processing threshold. An additional
$L_2$ penalty can be included straightforwardly as shown in Appendix~\ref{app:sparse_details}, yielding the usual
Elastic-Net generalization~\cite{zou2005regularization}. In the following subsection we retain this sparsity-promoting component and
extend the prior so that its shrinkage also depends on the inferred
excitatory or inhibitory identity of the presynaptic neuron.

\subsection{Dale-consistent connectivity}
\label{sec:Dale}

We next consider sign consistency among the outgoing interactions of each
presynaptic neuron. Motivated by Dale's principle \cite{strata1999dale,po2025inferring}, we associate neuron $j$
with a latent identity
\begin{equation}
    z_j\in\{-1,+1\},
\end{equation}
where $z_j=+1$ and $z_j=-1$ denote excitatory and inhibitory identities,
respectively. Conditional on these identities, the prior factorizes over
outgoing couplings,
\begin{equation}
    P(J,\vect z)
    =
    \prod_j
    P(z_j)
    \prod_{i\neq j}
    P(J_{ij}\mid z_j).
    \label{eq:Dale_factorization}
\end{equation}
The prior $P(z_j)$ can encode information about the expected neuronal
composition; when no such information is available we take it to be uniform.

A strict implementation of Dale's principle would constrain all outgoing
couplings of a neuron to have the same sign. Here, however, $J_{ij}$ represents
an effective interaction rather than a direct synaptic conductance, so exact
sign consistency need not be preserved under coarse graining, unobserved
inputs, or model mismatch. We therefore extend the sparse Laplace prior of
Sec.~\ref{sec:sparsity} by assigning an additional penalty to couplings
whose sign is inconsistent with the latent identity of their presynaptic
neuron.

Defining
$J_{ij}^{+}=\max(J_{ij},0)$ and
$J_{ij}^{-}=\max(-J_{ij},0)$, we take
\begin{align}
P(J_{ij}\mid z_j=+1)
&=
C\exp\left[
-\gamma_1 J_{ij}^{+}
-(\gamma_1+\beta)J_{ij}^{-}
\right],
\nonumber\\
P(J_{ij}\mid z_j=-1)
&=
C\exp\left[
-(\gamma_1+\beta)J_{ij}^{+}
-\gamma_1J_{ij}^{-}
\right],
\label{eq:dale_prior}
\end{align}
where $\beta\geq0$ controls the additional cost of a sign violation and
\begin{equation}
C=
\frac{\gamma_1(\gamma_1+\beta)}
     {2\gamma_1+\beta}.
\end{equation}
Thus $\gamma_1$ controls overall sparsity, while $\beta$ controls the
preference for Dale-consistent outgoing signs. For $\beta=0$ the prior reduces
to the symmetric Laplace prior of Sec.~\ref{sec:sparsity}.

To infer the neuronal identities jointly with the interaction matrix, we
extend the variational approximation of Sec.~\ref{sec:Inference} to include
a factor $Q_j(z_j)$ for each neuronal identity. Analogously to the
magnetizations $m_i(t)=Q_i^t(s_i(t)=1)$ introduced for the dynamical latent
states, we define
\begin{equation}
    \phi_j
    =
    Q_j(z_j=+1),
\end{equation}
so that $\phi_j$ is the variational probability that neuron $j$ has
an excitatory identity. For the prior in
Eq.~\eqref{eq:dale_prior}, its coordinate update is
\begin{equation}
\phi_j
=
\sigma\left[
\log
\frac{P(z_j=+1)}{P(z_j=-1)}
+
\beta\sum_{i\neq j}J_{ij}
\right].
\label{eq:phi_update}
\end{equation}
Thus the evidence for the neuronal identity is accumulated across all
outgoing effective interactions. The sparsity strength $\gamma_1$ does not
enter this update directly because it penalizes the two neuronal identities
symmetrically; only the additional sign-asymmetry $\beta$ contributes to their
relative probability. The variational derivation of this update is given in
Appendix~\ref{app:DaleDerivation}.

For fixed $\phi_j$, averaging the conditional log-prior over the neuronal identity
gives the effective penalties
\begin{equation}
    \widetilde{\gamma}_j^{+}
    =
    \gamma_1+\beta(1-\phi_j),
    \qquad
    \widetilde{\gamma}_j^{-}
    =
    \gamma_1+\beta\phi_j.
    \label{eq:dale_effective_penalties}
\end{equation}

After the same likelihood-gradient step used in Sec.~\ref{sec:sparsity},
$\widetilde J_{ij}=J_{ij}+\alpha G_{ij}$, the corresponding
proximal update is
\begin{equation}
    J_{ij}^{\rm new}
    =
    \begin{cases}
        \widetilde J_{ij}
        -
        \alpha\widetilde\gamma_j^{+},
        &
        \widetilde J_{ij}
        >
        \alpha\widetilde\gamma_j^{+},
        \\[4pt]
        0,
        &
        -\alpha\widetilde\gamma_j^{-}
        \leq
        \widetilde J_{ij}
        \leq
        \alpha\widetilde\gamma_j^{+},
        \\[4pt]
        \widetilde J_{ij}
        +
        \alpha\widetilde\gamma_j^{-},
        &
        \widetilde J_{ij}
        <
        -\alpha\widetilde\gamma_j^{-}.
    \end{cases}
    \label{eq:Dale_proximal}
\end{equation}
The common contribution $\gamma_1$ shrinks both signs and promotes sparsity,
whereas the $\beta$ contribution selectively increases the threshold for the
sign inconsistent with the current neuronal identity. Alternating the updates
of $\phi_j$ and $J$ therefore jointly promotes sparse connectivity and
Dale-consistent outgoing interactions while allowing finite-probability sign
violations.

\subsection{Validation on structured SpiKIsing networks}
\label{sec:structured_validation}

\begin{figure}
    \centering
    \begin{subfigure}[t]{0.48\linewidth}
        \centering
        \includegraphics[width=\linewidth]{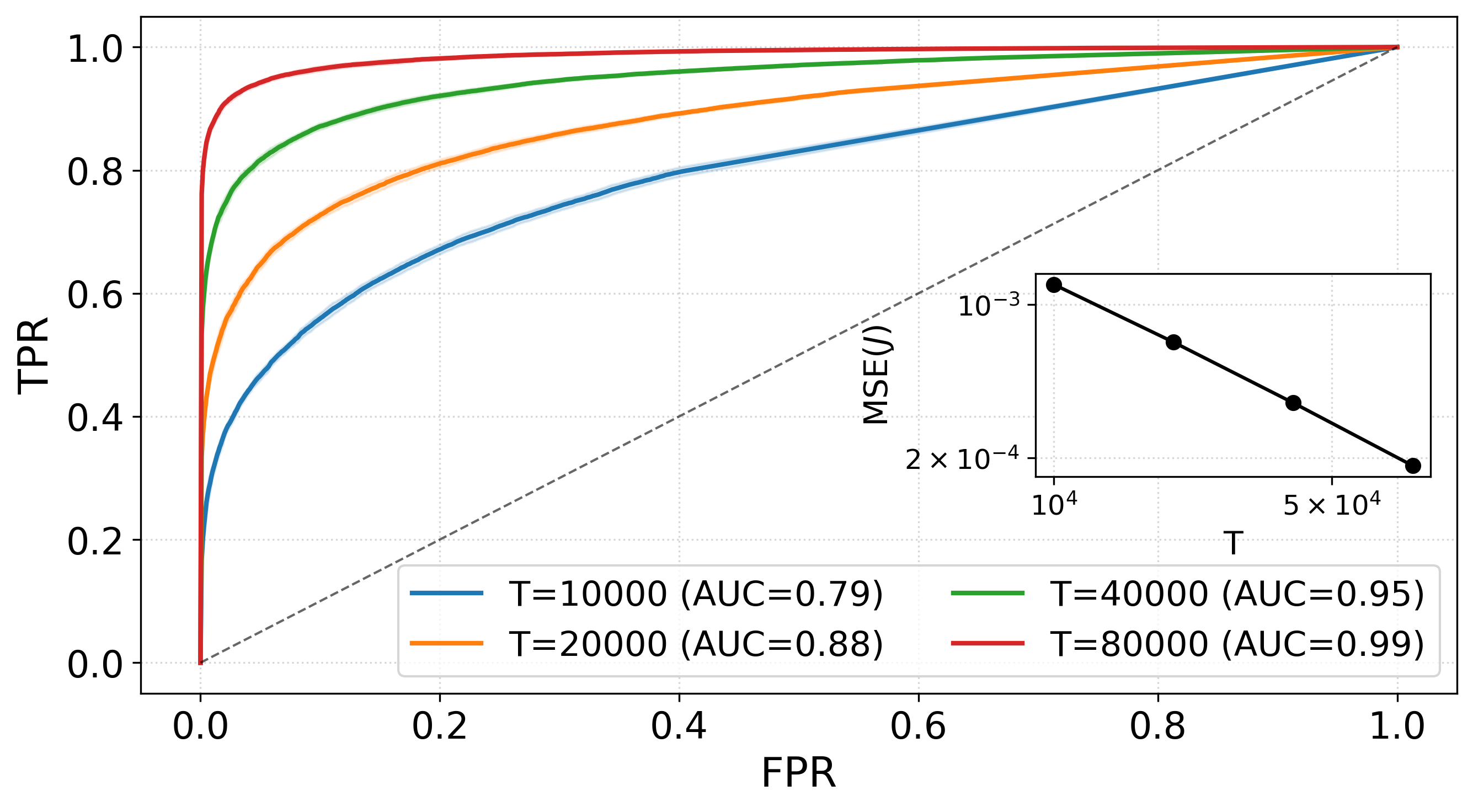}
        \caption{}
        \label{fig:Laplace}
    \end{subfigure}\hfill
    \begin{subfigure}[t]{0.48\linewidth}
        \centering
        \includegraphics[width=\linewidth]{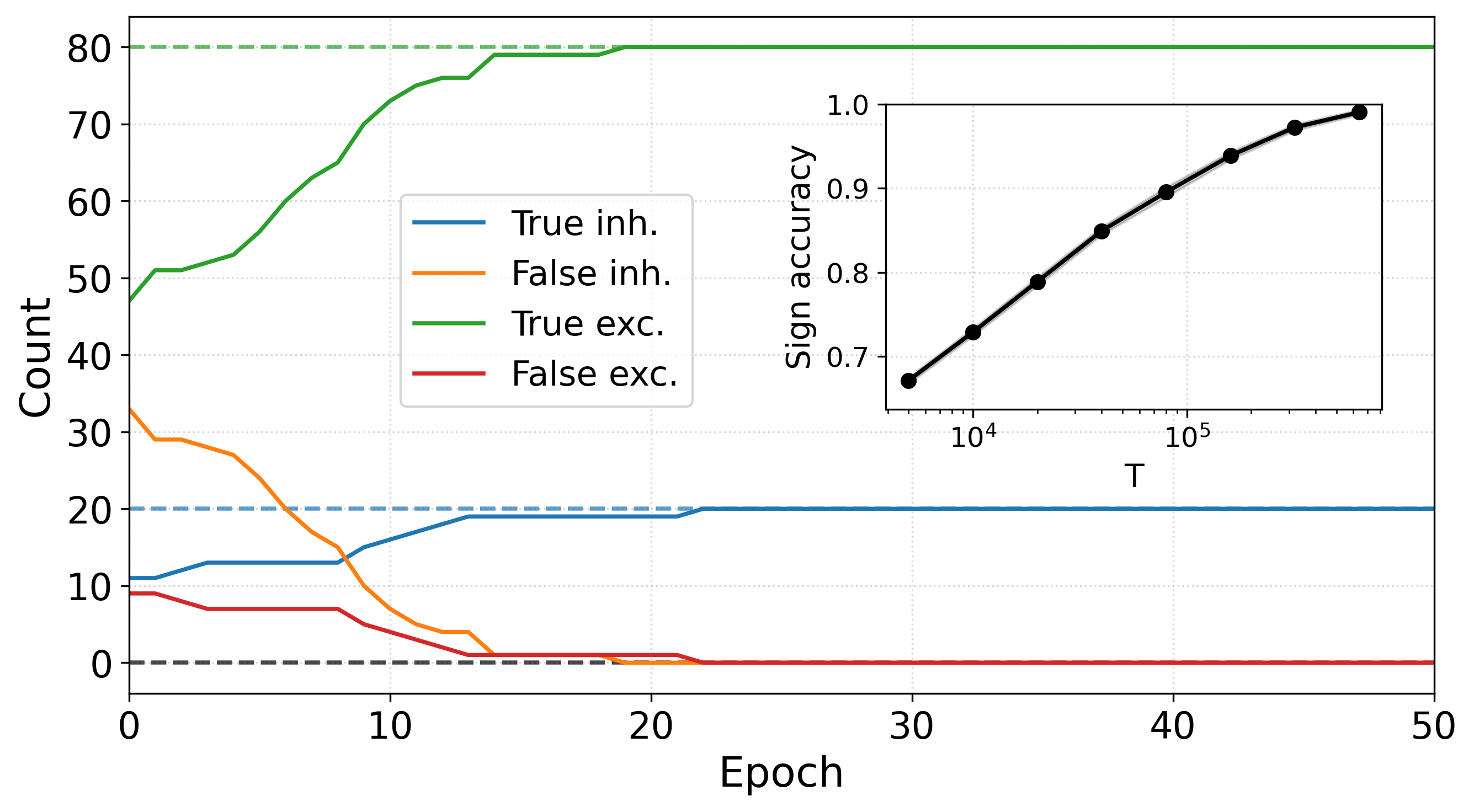}
        \caption{}
        \label{fig:DaleAccuracy}
    \end{subfigure}
     \caption{
    \textbf{Matched-model validation of sparse and Dale-structured MAP inference.}\\
    \textbf{(a)} \textbf{Sparse connectivity recovery.}
        Receiver Operating Characteristic (ROC) curves for classification of non-zero interactions at different
        recording lengths $T$. The inset shows the MSE of the inferred
        interaction matrix. \textbf{(b)} \textbf{Recovery of Dale-consistent structure.}
        Evolution of the neuronal-type classification during inference.
        The inset shows the fraction of ground-truth non-zero couplings whose
        sign is correctly inferred as a function of recording length $T$.
        Both experiments use networks of $N=100$ neurons. Further simulation and
    inference details are given in Appendices~\ref{app:sparse_details} and
    \ref{app:DaleDerivation}. 
    }
    \label{fig:StructuredPriors}
\end{figure}

We test the structured priors on spike trains generated from SpiKIsing
networks with known sparse or Dale-consistent connectivity. The experiment of
Fig.~\ref{fig:Laplace} uses the sparse Laplace prior alone, whereas
Fig.~\ref{fig:DaleAccuracy} combines the same $L_1$ sparsity prior with
the hierarchical Dale extension of Sec.~\ref{sec:Dale}. In both
experiments we use the forward-only variational update and the centred
parametrization of Appendix~\ref{app:centered-reparam}. To isolate the
effect of the prior on interaction recovery, the emission-model parameters,
including $\tau_i^{\rm abs}$, are fixed to their generating values. Details
of the sparse-network construction and regularization procedure are given in
Appendix~\ref{app:sparse_details}, while the derivation and numerical details
of the Dale-consistent inference are given in
Appendix~\ref{app:DaleDerivation}.

For the sparse-network test, the ground-truth interaction matrix contains an
explicit population of absent connections, and inference uses the $L_1$ prior
of Eq.~\eqref{eq:l1_prior}. We classify interaction support using
$|J_{ij}^{\rm inf}|$ as the score and vary the classification threshold to
construct the ROC curves in Fig.~\ref{fig:Laplace}. Recovery improves strongly
with recording length: the area under the ROC curve increases from
approximately $0.79$ at $T=10^4$ to $0.99$ at $T=8\times10^4$.
At the same time, the MSE of the inferred interaction matrix decreases,
showing improvement both in identification of the network support and in
quantitative estimation of the couplings.

We next test recovery of Dale-consistent structure in a network containing
$80\%$ excitatory and $20\%$ inhibitory neurons. The inference algorithm is
not supplied with this composition: we use an uninformative prior,
$P(z_j=+1)=P(z_j=-1)=1/2$. As shown in the main panel of
Fig.~\ref{fig:DaleAccuracy}, incorrect neuronal-type assignments decrease
during inference, while the inferred excitatory and inhibitory populations
approach their ground-truth values. The inset evaluates coupling-sign recovery
over ground-truth non-zero interactions. Sign accuracy increases with
recording length and approaches perfect recovery for the longest recordings
considered.

Together, these experiments illustrate the nested structure of the priors:
the Laplace term controls the support of the inferred interaction matrix,
while the hierarchical extension additionally couples the signs of outgoing
interactions through a latent neuronal identity.
Having established these properties in a controlled setting, we next test
SpiKIsing on spike trains generated by dynamics outside its own model class.

\section{Validation under conductance-based model mismatch}
\label{sec:LIF}

\begin{figure}
    \centering
    \includegraphics[width=0.9\linewidth]{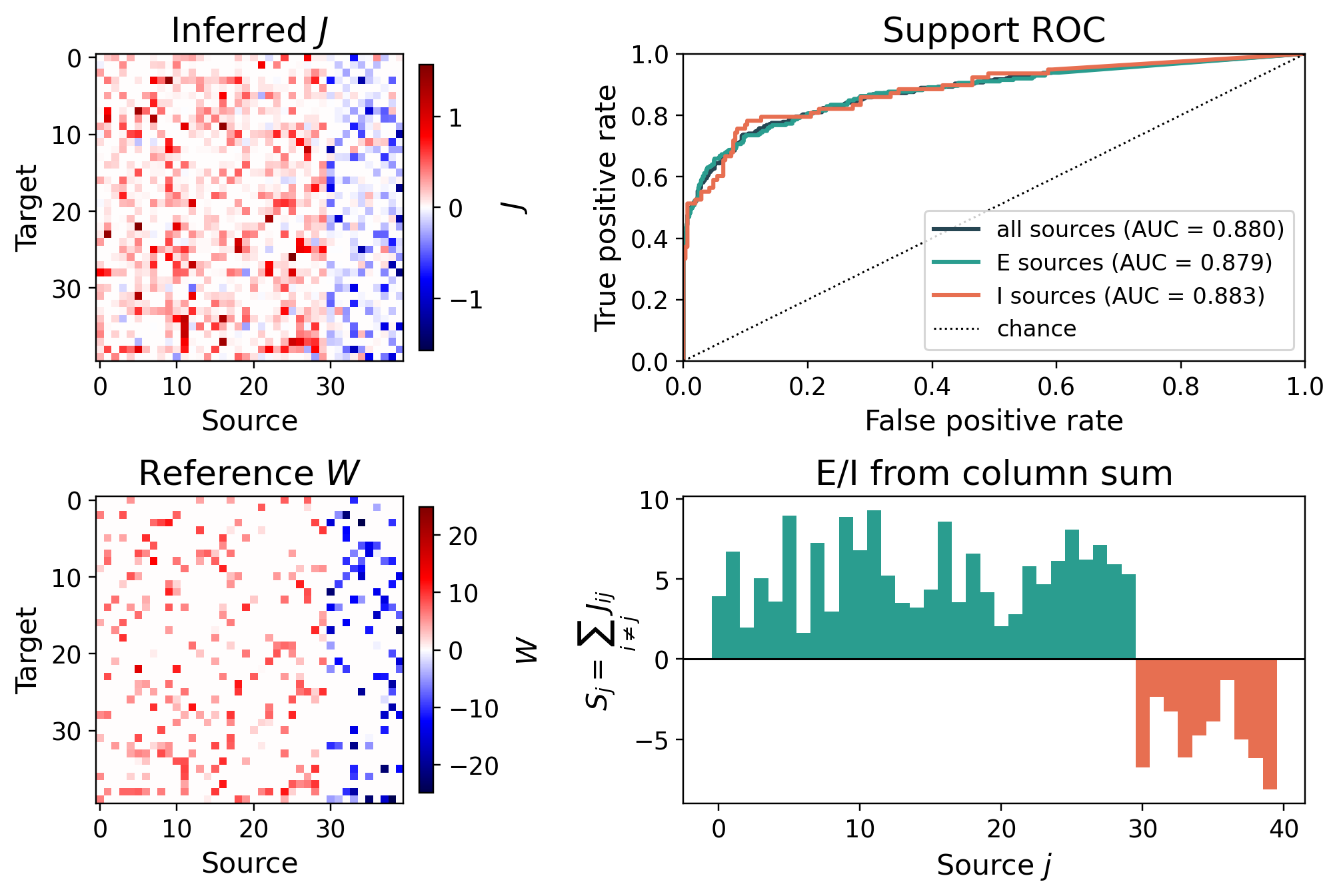}
    \caption{
    \textbf{Structural recovery in a conductance-based LIF network.}
    Results for $30$ excitatory and $10$ inhibitory neurons from a
    $98\,\mathrm{s}$ recording, inferred with latent interval
    $\Delta t=10\,\mathrm{ms}$.
    \textit{Upper left:} inferred effective interactions $J$.
    \textit{Lower left:} signed reference conductances $W$; rows denote targets
    and columns sources, with excitatory sources ordered first.
    \textit{Upper right:} ROC curves for recurrent-connection support using
    $|J_{ij}|$ as the score.
    \textit{Lower right:} outgoing interaction sums
    $S_j=\sum_{i\neq j}J_{ij}$, coloured by true neuronal type.
    }
    \label{fig:LIF_recovery}
\end{figure}

The preceding experiments test SpiKIsing under matched conditions, where the
latent dynamics and spike-generation process coincide with those assumed
during inference. A more stringent question is whether the inferred
interaction structure remains informative when neither assumption is
satisfied. We therefore generate spike trains from a recurrent
conductance-based leaky integrate-and-fire (LIF) network and ask whether
SpiKIsing can recover two structural properties that are known independently
from the simulator: the support of the recurrent connectivity and the
excitatory or inhibitory character of each presynaptic neuron.

We use a reduced recurrent network based on the conductance dynamics of
Ref.~\cite{renart2010asynchronous}; the full equations and numerical
parameters are given in Appendix~\ref{app:LIF_details}. The network contains
$30$ excitatory and $10$ inhibitory neurons with sparse recurrent
connectivity, together with independent external Poisson drive. All recurrent
neurons are observed, and external inputs are assigned privately to individual
neurons, so that we isolate dynamical model mismatch from
the additional complications of hidden recurrent units and shared external
input.

The mismatch is nevertheless substantial. Spikes are generated by threshold
crossings of continuously evolving membrane potentials, recurrent
interactions enter through voltage-dependent synaptic conductances, and
synaptic transmission has finite temporal dynamics and conduction delays.
None of these mechanisms is present in the SpiKIsing generative model.

We retain $98\,\mathrm{s}$ of simulated activity and infer the network using
the forward-only variational scheme with latent interval
$\Delta t=10\,\mathrm{ms}$. Specifically, we apply the combined sparse and Dale-structured MAP inference
of Sec.~\ref{sec:MAP} to spike trains generated by a recurrent
conductance-based leaky integrate-and-fire network. The emission parameters are inferred from
the spike trains, including the absolute refractory times, while the true
excitatory and inhibitory identities are withheld and assigned equal prior
probability. Further inference and hyperparameter details are given in
Appendix~\ref{app:LIF_details}.

The physical synaptic conductances and the inferred SpiKIsing couplings are
not expected to agree numerically: $J_{ij}$ measures an effective interaction
between latent states, whereas the simulator specifies a conductance acting
on a continuous membrane potential. We therefore compare structural rather
than metric properties. For evaluation, we define the signed reference matrix
\begin{equation}
    W_{ij}=z_j p_{ij}g_{ij},
    \qquad W_{ii}=0,
    \label{eq:LIF_reference}
\end{equation}
where $p_{ij}\in\{0,1\}$ indicates whether a recurrent connection from source
neuron $j$ to target neuron $i$ is present, $g_{ij}\geq0$ is its conductance
amplitude when present, and $z_j=+1$ ($z_j=-1$) for excitatory (inhibitory)
sources.
We then ask whether $|J_{ij}|$ discriminates present from absent recurrent
connections and whether the aggregate sign of the outgoing inferred
interactions identifies the presynaptic neuronal type.

Figure~\ref{fig:LIF_recovery} shows that the inferred matrix reproduces
part of the sparse structure of the reference network despite the dynamical
mismatch. Using $|J_{ij}|$ as a connection score gives an ROC area of
$0.88$ over all off-diagonal pairs. The corresponding values are $0.879$
and $0.883$ when evaluation is restricted to excitatory and inhibitory
sources, respectively, indicating comparable support discrimination for the
two presynaptic populations.

The same inferred interactions also contain a clear population-level sign
structure. For neuron $j$, define the total outgoing effective interaction
\begin{equation}
    S_j=\sum_{i\neq j}J_{ij}.
    \label{eq:LIF_source_score}
\end{equation}
For the uniform identity prior used here, Eq.~\eqref{eq:phi_update} reduces to
\begin{equation}
    \phi_j=\sigma(\beta S_j).
\end{equation}
Since $\beta>0$, thresholding $\phi_j$ at $1/2$ is exactly equivalent to
classifying the presynaptic identity according to the sign of $S_j$.
As shown in Fig.~\ref{fig:LIF_recovery}, all $30$ excitatory neurons have
$S_j>0$, whereas all $10$ inhibitory neurons have $S_j<0$. Neuronal type is
therefore classified correctly for all $40$ neurons in this realization,
even though individual inferred couplings need not obey exact sign
consistency.

\section{\label{sec:Conclusions}Conclusions and outlook}

We have introduced SpiKIsing, a latent-variable framework for inferring
directed effective interactions from neuronal spike trains. Interacting
binary states describe the collective network dynamics, while
history-dependent point processes describe how those states generate
spikes in continuous time. This separation allows the timing of
individual spikes, including refractory and recovery effects, to
inform inference without identifying each spike directly with an
Ising spin. In the resulting variational inference scheme, the
point-process likelihood enters the latent-state update as an
observation field. Sparse and Dale-structured priors then allow
additional information about network organization to be incorporated
into the coupling estimates.

In matched-model experiments, interaction reconstruction improved with
recording length, and joint inference recovered the latent parameters
together with the emission parameters. On networks with known sparse and Dale-consistent structure, the resulting
MAP estimates recovered connection support and outgoing signs.
The conductance-based LIF test provided a more stringent check under
model mismatch: in the realization studied, inferred coupling
magnitudes discriminated recurrent connections with an ROC area of
$0.88$, and the outgoing interaction sums identified the neuronal
type of all $40$ recurrent neurons. These results show that the
inferred interactions can retain useful structural information even
when the spike-generating dynamics differs substantially from the
model used for inference.

The inferred couplings should nevertheless be interpreted as \textit{effective} interactions between latent states, rather than direct
measurements of synaptic conductances. Accordingly, the LIF
experiment evaluates connection support and presynaptic sign, not
numerical agreement between couplings and conductances. Moreover,
although observed spike times remain continuous, the present latent
dynamics still evolves on a chosen discrete time step. The effect of
that choice, and the performance of the method on experimental
recordings, remain to be assessed.

The present inference scheme also trades accuracy for tractability.
Replacing the fluctuating latent field by its mean means that the
optimized nMF objective is not guaranteed to remain a strict
variational lower bound; the forward-only update introduces a further
approximation by omitting the backward reaction term. A natural next
step is to investigate higher-order dynamical mean-field or
TAP-like corrections~\cite{roudi2011dynamical,bachschmid2016variational},
particularly at stronger coupling, while measuring their additional
computational cost and comparing them with the full forward--backward
update.

A second direction is to formulate the latent network dynamics in
continuous time, avoiding an imposed clock for state transitions
while retaining the point-process description of the observed
spikes~\cite{zeng2013maximum,donner2017inverse}. The emission model
could also be made more flexible. In particular, the rational
post-spike recovery function used here is a phenomenological choice;
a constrained basis expansion or neural-network parametrization of
$\rho_i$ could capture more varied recovery profiles. Such
flexibility would require regularization and evaluation on held-out
spike trains, since an overly expressive emission model might fit
noise or absorb temporal structure that should instead inform the
latent dynamics and couplings.

Additionally, the LIF benchmark includes unobserved external Poisson
sources, but these provide private input to individual neurons:
all recurrent neurons are observed, and neither hidden recurrent
units nor shared external drive are included. Both complications
may alter dependencies among recorded neurons and the relationship
between inferred effective interactions and physical connections
\cite{brinkman2018predicting}. Extending SpiKIsing to account for
hidden recurrent units and correlated input is therefore an
important step toward experimental applications. Methods for
kinetic Ising models with hidden spins
\cite{dunn2013learning,battistin2015belief} offer possible starting
points, although genuinely unrecorded neurons must be distinguished
from the latent states already associated with recorded neurons.
Together, these extensions would help establish when spike-time-based
latent network inference can reliably characterize the structure of
partially observed neural circuits.

Finally, the aim of the model and inference technique is to infer effective neuronal networks from \textit{in vivo} neuronal culture recordings, which introduces additional challenges. Applying our work to real neuronal networks is underway.

\section*{Code availability}
The core SpiKIsing inference implementation and some illustrative examples are
available at \url{https://github.com/DavideGhioEPFL/SpiKIsing/}

\section*{Acknowledgments and Statement of AI usage}
Support from the UK Multidisciplinary Centre for Neuromorphic Computing (UKRI982) is gratefully acknowledged. We have used  Generative-AI models at different stages of the project: GitHub Copilot (Microsoft) has been used extensively as coding assistant and GPT-5.6 Sol (OpenAI) has helped with revising and editing the text. The authors take full responsibility for the final contents of the manuscript.

\newpage
\appendix

\section{Emission-model quantities and parameter updates}
\label{app:DerC}

In this appendix we collect the analytical results used in the M-step for the
emission model. We derive the closed-form updates for the active- and
inactive-state firing rates, evaluate the recovery-weighted exposure $A_i^t$
for the kernel of Eq.~\eqref{eq:recovery}, and give the derivatives required
to optimize the recovery parameter $c_i$. We also describe the profile
optimization used to infer the absolute refractory time
$\tau_i^{\rm abs}$ when it is treated as an unknown parameter. 

\subsection{Active- and inactive-state firing rates}

For fixed magnetizations and recovery parameters, the dependence of the
variational objective on $\lambda_i^+$ and $\lambda_i^-$ is contained in
the observation term,
\begin{align}
    \mathcal L_{{\rm obs},i}
    =
    \sum_t
    \Big[
        &(1-m_i(t))
        \left(
            n_i^t\log\lambda_i^-
            -
            \lambda_i^-E_i^t
        \right)
        \nonumber\\
        &+
        m_i(t)
        \left(
            n_i^t\log\lambda_i^+
            +
            C_i^t
            -
            \lambda_i^+A_i^t
        \right)
    \Big].
    \label{eq:Lobs_i_app}
\end{align}
Differentiating with respect to the two firing rates gives
\begin{align}
    \frac{\partial\mathcal F_{\rm nMF}}
         {\partial\lambda_i^+}
    &=
    \sum_t
    m_i(t)
    \left(
        \frac{n_i^t}{\lambda_i^+}
        -
        A_i^t
    \right),
    \label{eq:grad_lam_plus_app}
    \\
    \frac{\partial\mathcal F_{\rm nMF}}
         {\partial\lambda_i^-}
    &=
    \sum_t
    (1-m_i(t))
    \left(
        \frac{n_i^t}{\lambda_i^-}
        -
        E_i^t
    \right).
    \label{eq:grad_lam_minus_app}
\end{align}
Setting these derivatives to zero yields the closed-form updates
given in Eqs.~\eqref{eq:update_lam_plus} and
\eqref{eq:update_lam_minus}.

\subsection{Recovery-weighted exposure}

For the recovery function introduced in Eq.~\eqref{eq:recovery},
\begin{equation*}
    \rho_i(\delta)
    =
    \frac{\delta}{\delta+c_i},
    \qquad
    \delta>0,
\end{equation*}
the recovery-weighted exposure $A_i^t$ can be evaluated analytically.

To evaluate $A_i^t$, we partition the latent interval $[t,t+1)$ at every
observed spike of neuron $i$. Within each resulting sub-interval
$[u_a,u_b)$, the most recent spike is fixed; denote its time by
$\widetilde{\tau}_i$. We define the effective integration limits
\begin{equation}
    a
    =
    \max\left(
        0,
        u_a-\widetilde{\tau}_i-\tau_i^{\rm abs}
    \right),
    \qquad
    b
    =
    \max\left(
        0,
        u_b-\widetilde{\tau}_i-\tau_i^{\rm abs}
    \right).
    \label{eq:ab_limits}
\end{equation}
The contribution of this sub-interval to the active-state exposure is
\begin{align}
    A_{i,[a,b]}
    &=
    \int_a^b
    \frac{\delta}{\delta+c_i}\,d\delta
    \nonumber\\
    &=
    (b-a)
    -
    c_i
    \log
    \left(
        \frac{b+c_i}{a+c_i}
    \right),
    \label{eq:A_segment}
\end{align}
while the corresponding contribution to the inactive-state exposure is
\begin{equation*}
    E_{i,[a,b]}=b-a.
\end{equation*}
Summing Eq.~\eqref{eq:A_segment} over all sub-intervals contained in
$[t,t+1)$ gives $A_i^t$. Before the first observed spike, the neuron is
assumed to be fully recovered, as described in Sec.~\ref{sec:Model}; the
corresponding segment contributes its full available duration to $A_i^t$
and carries no dependence on $c_i$.

For fixed $\tau_i^{\rm abs}$, the integration limits are independent of
$c_i$, and differentiation of Eq.~\eqref{eq:A_segment} gives
\begin{equation}
    \frac{\partial A_{i,[a,b]}}{\partial c_i}
    =
    -
    \log
    \left(
        \frac{b+c_i}{a+c_i}
    \right)
    -
    c_i
    \left(
        \frac{1}{b+c_i}
        -
        \frac{1}{a+c_i}
    \right).
    \label{eq:dA_segment}
\end{equation}
Therefore,
\begin{equation}
    \frac{\partial A_i^t}{\partial c_i}
    =
    \sum_{[a,b]\subset[t,t+1)}
    \frac{\partial A_{i,[a,b]}}{\partial c_i}.
    \label{eq:dA_total}
\end{equation}

\subsection{Recovery contribution and update of \texorpdfstring{$c_i$}{c}}

For an observed spike at time $\tau_{i,k}$ with $k\geq2$, define
\begin{equation}
    \Delta_{i,k}
    =
    \tau_{i,k}
    -
    \tau_{i,k-1}
    -
    \tau_i^{\rm abs}.
\end{equation}
For an admissible spike train, $\Delta_{i,k}>0$, and
\begin{equation*}
    \log\rho_i(\Delta_{i,k})
    =
    \log\Delta_{i,k}
    -
    \log(\Delta_{i,k}+c_i).
\end{equation*}
Hence
\begin{equation*}
    \frac{\partial}{\partial c_i}
    \log\rho_i(\Delta_{i,k})
    =
    -
    \frac{1}{\Delta_{i,k}+c_i},
\end{equation*}
and the derivative of the interval spike contribution is
\begin{equation}
    \frac{\partial C_i^t}{\partial c_i}
    =
    -
    \sum_{\substack{\tau_{i,k}\in[t,t+1)\\k\geq2}}
    \frac{1}{\Delta_{i,k}+c_i}.
    \label{eq:dC_dc}
\end{equation}
The restriction $k\geq2$ reflects the fully recovered initial condition
used for the first observed spike.

The terms of $\mathcal F_{\rm nMF}$ that depend explicitly on $c_i$ are
\begin{equation}
    \mathcal F_i^{(c)}
    =
    \sum_{t=0}^{T-1}
    m_i(t)
    \left[
        C_i^t-\lambda_i^+A_i^t
    \right].
    \label{eq:F_c}
\end{equation}
The gradient used in the M-step is therefore
\begin{equation}
    \frac{\partial\mathcal F_{\rm nMF}}{\partial c_i}
    =
    \sum_{t=0}^{T-1}
    m_i(t)
    \left[
        \frac{\partial C_i^t}{\partial c_i}
        -
        \lambda_i^+
        \frac{\partial A_i^t}{\partial c_i}
    \right].
    \label{eq:grad_c_app}
\end{equation}
Substitution of Eqs.~\eqref{eq:dA_total} and \eqref{eq:dC_dc} gives the
analytical gradient used to update $c_i$ by gradient ascent.

\subsection{Profile estimation of the absolute refractory time}
When $\tau_i^{\rm abs}$ is not fixed independently, we estimate it by
one-dimensional profile maximization of the observation likelihood. The
parameter affects both the support of the point process and the exposure terms
$E_i^t$, $A_i^t$, and $C_i^t$, so a direct numerical optimization is more
convenient than an analytical gradient update.

The refractory time must be shorter than the smallest observed inter-spike
interval,
\begin{equation}
    d_i^{\rm min}
    =
    \min_{k\geq2}
    \left(
        \tau_{i,k}-\tau_{i,k-1}
    \right).
\end{equation}
We therefore optimize over
\begin{equation}
    0
    \leq
    \tau_i^{\rm abs}
    \leq
    d_i^{\rm min}-\epsilon_i,
    \qquad
    \epsilon_i>0,
    \label{eq:tau_abs_bounds}
\end{equation}
where $\epsilon_i$ keeps all post-refractory intervals strictly positive.
For each trial value of $\tau_i^{\rm abs}$, the quantities $E_i^t$, $A_i^t$,
and $C_i^t$ are recomputed and inserted into the observation likelihood.

\section{Derivation of the variational bound}
\label{app:Var}

For completeness, we derive here the variational bound used in
Sec.~\ref{sec:Inference}. Denote the complete set of model parameters by
$\theta=(\theta_{\rm lat},\theta_{\rm obs})$. Starting from the marginal
likelihood,
\begin{equation}
    P(\vect{\tau}\mid\theta)
    =
    \sum_{\vect{s}_{0:T-1}}
    P(\vect{\tau},\vect{s}_{0:T-1}\mid\theta),
\end{equation}
we introduce an arbitrary normalized distribution
$Q(\vect{s}_{0:T-1})$ and write
\begin{align}
    \mathcal L
    &=
    \log P(\vect{\tau}\mid\theta)
    \nonumber\\
    &=
    \log
    \sum_{\vect{s}_{0:T-1}}
    Q(\vect{s}_{0:T-1})
    \frac{
        P(\vect{\tau},\vect{s}_{0:T-1}\mid\theta)
    }{
        Q(\vect{s}_{0:T-1})
    }.
\end{align}

Using Jensen's inequality,
\begin{align}
    \mathcal L
    &\geq
    \sum_{\vect{s}_{0:T-1}}
    Q(\vect{s}_{0:T-1})
    \log
    \frac{
        P(\vect{\tau},\vect{s}_{0:T-1}\mid\theta)
    }{
        Q(\vect{s}_{0:T-1})
    }
    \nonumber\\
    &\equiv
    \mathcal F[Q].
    \label{eq:ELBO_app}
\end{align}

Equivalently,
\begin{equation}
    \mathcal L-\mathcal F[Q]
    =
    D_{\rm KL}
    \left[
        Q(\vect{s}_{0:T-1})
        \,\Vert\,
        P(\vect{s}_{0:T-1}\mid\vect{\tau},\theta)
    \right]
    \geq0.
    \label{eq:KL_ELBO}
\end{equation}
The bound is saturated when $Q$ coincides with the exact posterior over the
latent trajectory.

Using the factorization of the SpiKIsing generative model,
\begin{equation}
    P(\vect{\tau},\vect{s}_{0:T-1}\mid\theta)
    =
    P(\vect{\tau}\mid\vect{s}_{0:T-1},\theta_{\rm obs})
    P(\vect{s}_{0:T-1}\mid\theta_{\rm lat}),
\end{equation}
the variational free energy becomes
\begin{equation}
    \mathcal F[Q]
    =
    \mathbb E_Q
    \left[
        \log P(\vect{\tau}\mid\vect{s}_{0:T-1},\theta_{\rm obs})
    \right]
    +
    \mathbb E_Q
    \left[
        \log P(\vect{s}_{0:T-1}\mid\theta_{\rm lat})
    \right]
    +
    \mathcal S_Q,
\end{equation}
which gives the decomposition
$\mathcal F=\mathcal L_{\rm obs}+\mathcal L_{\rm lat}+\mathcal S_Q$
used in the main text.

We stress that the inequality above applies to the exact variational
functional $\mathcal F[Q]$. The subsequent replacement of the fluctuating
kinetic-Ising field by its mean, leading to $\mathcal F_{\rm nMF}$ in
Eq.~\eqref{eq:F_nMF}, constitutes an additional approximation and does not in
general preserve the rigorous lower-bound property.

\section{Comparison of forward-only and forward--backward schemes\label{app:FBvsB}}

In Sec.~\ref{sec:Inference} we introduced two E-step updates for the
variational magnetizations. The full forward--backward scheme retains the
reaction term
\begin{equation*}
    R_i(t)=\sum_k J_{ki}\left[m_k(t+1)-\sigma(\eta_k(t+1))\right],
\end{equation*}
and iterates Eq.~\eqref{eq:mean_field_exact} to a stationary point of
$\mathcal F_{\rm nMF}$. The forward-only approximation instead neglects this
backward contribution and updates the magnetizations causally according to
Eq.~\eqref{eq:mean_field_forward}. The latter is computationally simpler, since each
E-step requires only a single forward pass, but it is not in general a
stationary solution of $\mathcal F_{\rm nMF}$.

To assess the practical effect of this approximation, we compare the two
schemes on a representative matched-model dataset generated from SpiKIsing
with the same rational recovery kernel as used throughout the main matched
experiments. We use $N=100$ neurons and recording length $T=50000$, with the
same generative setting as in Sec.~\ref{sec:matched_validation}. Both inference
runs are initialized identically and use the same M-step updates; they differ
only in the treatment of the E-step.

\begin{figure}
    \centering
    \includegraphics[width=0.95\linewidth]{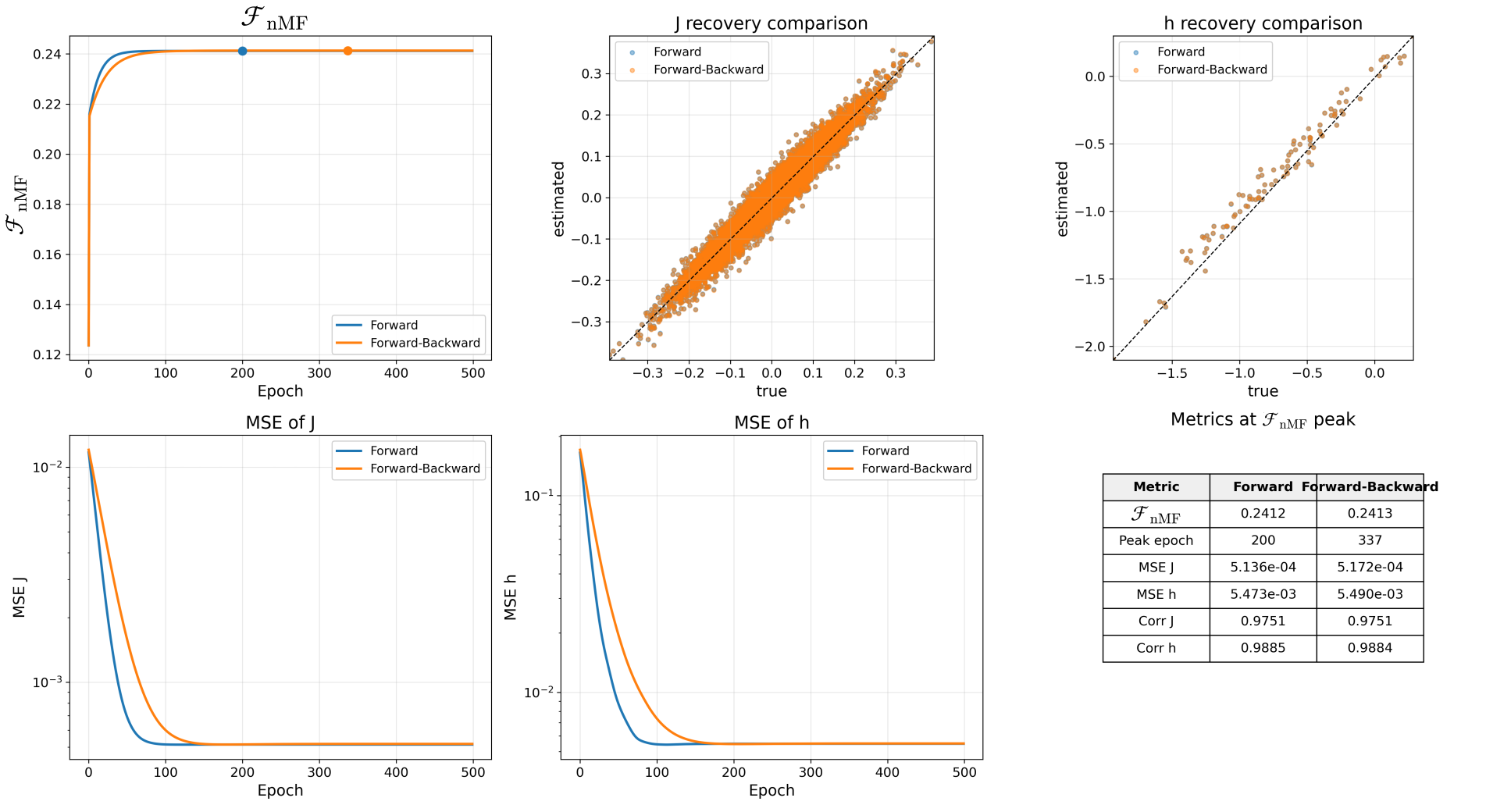}
    \caption{
    \textbf{Comparison of forward-only and forward--backward variational
    updates in a matched SpiKIsing experiment.}
    Results are shown for a representative dataset with $N=100$ neurons and
    recording length $T=50000$.
    \textit{Top left:} evolution of the nMF objective
    $\mathcal F_{\rm nMF}$ during inference; filled markers indicate the epoch
    at which each scheme reaches its maximum value.
    \textit{Top middle and top right:} inferred versus ground-truth couplings
    $J_{ij}$ and local fields $h_i$, respectively, evaluated at the epoch
    maximizing $\mathcal F_{\rm nMF}$ for each method.
    \textit{Bottom left and bottom middle:} mean-squared errors of the inferred
    couplings and local fields as functions of epoch.
    \textit{Bottom right:} summary metrics at the maximum of
    $\mathcal F_{\rm nMF}$ for each scheme.
    }
    \label{fig:FBvsF}
\end{figure}

Figure~\ref{fig:FBvsF} shows that the two schemes behave very similarly.
The forward--backward update achieves a marginally larger maximum objective but this does not translate into a meaningful improvement in
parameter recovery.

At the respective objective maxima, the two schemes yield almost identical
reconstructions of both the interaction matrix and the local fields. 
The scatter plots in Fig.~\ref{fig:FBvsF} confirm that the inferred
parameters from the two schemes overlap almost perfectly.

The optimization trajectories further show that the forward-only
approximation reaches its plateau in fewer outer epochs in this example,
whereas the forward--backward scheme converges more slowly while requiring a
more expensive E-step. Since the gain in objective value is negligible and
the recovered parameters are essentially unchanged, we adopt the forward-only
update in the numerical experiments reported in the main text and in the
following appendices.

\section{Centred parametrization for stable optimization}
\label{app:centered-reparam}

The $0/1$ representation of the latent spins is convenient for the
emission model, but it introduces a numerical issue in the M-step.
Because the presynaptic magnetizations $m_j(t)$ generally have non-zero
temporal means, changes in the row sums of $J$ can be partially compensated
by changes in the local fields $h_i$. This produces strong correlations
between the intercept and coupling parameters and can lead to poorly
conditioned optimization.

We therefore perform the M-step in an equivalent centred parametrization.
For the current variational trajectory, define the temporal mean of each
presynaptic variable as
\begin{equation}
    \mu_j
    =
    \frac{1}{T-1}
    \sum_{t=1}^{T-1}m_j(t-1)
    =
    \frac{1}{T-1}
    \sum_{t=0}^{T-2}m_j(t).
    \label{eq:mu_center}
\end{equation}
We then introduce the centred predictors
\begin{equation}
    x_j(t-1)
    =
    m_j(t-1)-\mu_j
\end{equation}
and the reparametrized local field
\begin{equation}
    b_i
    =
    h_i+\sum_jJ_{ij}\mu_j.
    \label{eq:b_center}
\end{equation}

The mean field can consequently be written as
\begin{equation}
    \eta_i(t)
    =
    b_i
    +
    \sum_jJ_{ij}x_j(t-1).
    \label{eq:eta_centered}
\end{equation}
For fixed $\vect{\mu}$ this expression is algebraically identical to
Eq.~\eqref{eq:eta_def}; the transformation therefore does not modify the
model or introduce an additional approximation.

During the M-step we hold $\vect{\mu}$ fixed and optimize the parameters
$(J,\vect{b})$. Defining the residual
\begin{equation}
    r_i(t)
    =
    m_i(t)-\sigma(\eta_i(t)),
\end{equation}
the centred coupling gradient at fixed $\vect b$ is
\begin{equation}
    G_{ij}
    \equiv
    \left.
    \frac{\partial\mathcal F_{\rm nMF}}{\partial J_{ij}}
    \right|_{\vect b}
    =
    \sum_{t=1}^{T-1}
    x_j(t-1)\,r_i(t).
    \label{eq:grad_J_centered}
\end{equation}
The derivative with respect to the centred intercept is
\begin{equation}
    \frac{\partial\mathcal F_{\rm nMF}}{\partial b_i}
    =
    \sum_{t=1}^{T-1}r_i(t),
\end{equation}
so that the same one-dimensional root condition used for $h_i$ can be
written as
\begin{equation}
    \sum_{t=1}^{T-1}
    \sigma
    \left(
        b_i+\sum_jJ_{ij}x_j(t-1)
    \right)
    =
    \sum_{t=1}^{T-1}m_i(t).
    \label{eq:b_root}
\end{equation}

After the M-step, the original local field is recovered exactly through
\begin{equation}
    h_i
    =
    b_i-\sum_jJ_{ij}\mu_j.
    \label{eq:h_from_b}
\end{equation}

The temporal means $\mu_j$ are recomputed from the current variational
trajectory at the beginning of each M-step and kept fixed during the
parameter update. In this way the centred variables describe fluctuations
around the average presynaptic activity, while $b_i$ absorbs the mean input.
This substantially improves the numerical conditioning of the optimization
without changing the likelihood represented by the model.

In the MAP setting considered in Sec.~\ref{sec:MAP}, priors acting only on
$J$ are unaffected by this change of coordinates. If a prior on $h_i$ were
introduced, it would instead have to be transformed according to
\begin{equation}
    P(h_i)
    =
    P\left(
        b_i-\sum_jJ_{ij}\mu_j
    \right),
\end{equation}
and would therefore couple the centred intercept and the corresponding row
of $J$.

\section{Sparse-prior inference and validation details}
\label{app:sparse_details}

In Sec.~\ref{sec:sparsity} we have used an $L_1$ prior to promote sparse estimates
of the interaction matrix. Here we give the corresponding Elastic-Net
generalization and summarize the simulation and evaluation procedure used
for the matched sparse-network experiment of Fig.~\ref{fig:Laplace}.

\subsection{Elastic-Net extension}

A more general sparse prior combines $L_1$ regularization with an $L_2$
penalty,
\begin{equation}
    P(J)
    \propto
    \exp\left[
        -\gamma_1\sum_{i\neq j}|J_{ij}|
        -
        \gamma_2\sum_{i\neq j}J_{ij}^2
    \right],
    \label{eq:elastic_prior}
\end{equation}
where $\gamma_1\geq0$ controls the sparsity-promoting contribution and
$\gamma_2\geq0$ provides additional shrinkage of the couplings. The pure
$L_1$ prior of Sec.~\ref{sec:sparsity} is recovered for $\gamma_2=0$.

At fixed magnetizations, the smooth contribution to the coupling update is
evaluated using the centred parametrization of
Appendix~\ref{app:centered-reparam}. Including the differentiable $L_2$
contribution gives
\begin{equation}
    G_{ij}^{\rm smooth}
    =
    G_{ij}
    -
    2\gamma_2J_{ij},
    \label{eq:grad_J_l2}
\end{equation}
where $G_{ij}$ is given by Eq.~\eqref{eq:grad_J_centered}.
A gradient-ascent step on the smooth part gives
\begin{equation}
    \widetilde J_{ij}
    =
    J_{ij}
    +
    \alpha G_{ij}^{\rm smooth},
    \label{eq:elastic_gradient_step}
\end{equation}
where $\alpha$ is the learning rate. The remaining non-differentiable
$L_1$ contribution is treated by soft thresholding,
\begin{equation}
    J_{ij}^{\rm new}
    =
    \operatorname{sign}(\widetilde J_{ij})
    \max\left(
        0,
        |\widetilde J_{ij}|-\alpha\gamma_1
    \right).
    \label{eq:elastic_proximal}
\end{equation}
For $\gamma_2=0$, this reduces to the update used in
Sec.~\ref{sec:sparsity}.

\subsection{Matched sparse-network experiment}

For the experiment shown in Fig.~\ref{fig:Laplace}, the off-diagonal
ground-truth couplings are drawn independently from a Laplace distribution,
\begin{equation}
    P_{\rm gen}(J_{ij})
    =
    \frac{\gamma_{\rm gen}}{2}
    \exp\left(
        -\gamma_{\rm gen}|J_{ij}|
    \right),
    \qquad
    \gamma_{\rm gen}=20.
    \label{eq:sparse_generation}
\end{equation}
To generate an interaction matrix with an explicit population of absent
connections, couplings satisfying
\begin{equation*}
    |J_{ij}|<J_{\rm min},
    \qquad
    J_{\rm min}=0.035,
\end{equation*}
are set to zero. The remaining couplings retain both positive and negative
signs.

Inference is performed with the pure $L_1$ limit of
Eq.~\eqref{eq:elastic_prior}, using
$\gamma_1=20$ and $\gamma_2=0$. As in the other structured matched-model
experiments, we use networks of $N=100$ neurons, the forward-only
variational update, and the centred M-step parametrization. The
emission-model parameters are fixed to their generating values.

To evaluate recovery of the interaction support, we use
$|J_{ij}^{\rm inf}|$ as a classification score and vary the threshold applied
to this quantity to construct the receiver-operating-characteristic curve
over all off-diagonal interactions. The corresponding area under the curve
quantifies discrimination between ground-truth zero and non-zero couplings.
Quantitative reconstruction of the coupling values is evaluated using the
mean-squared error defined in Eq.~\eqref{eq:MSE_J}.

\section{Variational derivation of the Dale-consistent prior}
\label{app:DaleDerivation}

In Sec.~\ref{sec:Dale} we extend the sparsity-promoting Laplace prior by
introducing a latent neuronal identity
$z_j\in\{-1,+1\}$ for each presynaptic neuron. The baseline $L_1$ penalty of
strength $\gamma_1$ acts on all couplings, while an additional penalty
$\beta$ is applied when the coupling sign is inconsistent with $z_j$. Here we derive the variational
update for the neuronal-type probabilities and its specialization to the
sign-dependent Laplace prior used in the main text.

The joint prior over the neuronal identities and interaction matrix is
\begin{equation}
    P(J,\vect z)
    =
    \prod_j
    P(z_j)
    \prod_{i\neq j}
    P(J_{ij}\mid z_j).
    \label{eq:Dale_joint_prior_app}
\end{equation}
We extend the mean-field variational distribution of
Sec.~\ref{sec:Inference} to include the neuronal identities,
\begin{equation}
    Q(\vect{s}_{0:T-1},\vect z)
    =
    \prod_{i,t}Q_i^t(s_i(t))
    \prod_jQ_j(z_j),
    \label{eq:Q_Dale}
\end{equation}
where the factors $Q_i^t$ are unchanged from the dynamical inference. We
parameterize the additional factors as
\begin{equation}
    Q_j(z_j)
    =
    \begin{cases}
        \phi_j, & z_j=+1,\\
        1-\phi_j, & z_j=-1,
    \end{cases}
    \label{eq:Q_z}
\end{equation}
so that $\phi_j$ is the variational probability that neuron $j$ has an
excitatory identity.

Up to terms independent of $\vect z$, the corresponding approximate
variational MAP objective is
\begin{align}
    \widetilde{\mathcal F}_{\rm nMF}
    =
    \mathcal F_{\rm nMF}
    &+
    \sum_{i\neq j}
    \left[
        \phi_j\log P(J_{ij}\mid+1)
        +
        (1-\phi_j)\log P(J_{ij}\mid-1)
    \right]
    \nonumber\\
    &+
    \sum_j
    \left[
        \phi_j\log P(z_j=+1)
        +
        (1-\phi_j)\log P(z_j=-1)
    \right]
    \nonumber\\
    &-
    \sum_j
    \left[
        \phi_j\log\phi_j
        +
        (1-\phi_j)\log(1-\phi_j)
    \right].
    \label{eq:elbo_dale}
\end{align}
The first additional term is the expected log-prior on the couplings, the
second is the prior contribution of the neuronal identities, and the last
term is the entropy of their variational distribution.

For fixed $J$, the objective factorizes over presynaptic neurons. Setting
the derivative with respect to $\phi_j$ to zero gives
\begin{equation}
    \log\frac{\phi_j}{1-\phi_j}
    =
    \log
    \frac{P(z_j=+1)}{P(z_j=-1)}
    +
    \sum_{i\neq j}
    \log
    \frac{
        P(J_{ij}\mid z_j=+1)
    }{
        P(J_{ij}\mid z_j=-1)
    }.
    \label{eq:phi_stationarity_app}
\end{equation}
Equivalently,
\begin{equation}
    \phi_j
    =
    \sigma
    \left[
        \log
        \frac{P(z_j=+1)}{P(z_j=-1)}
        +
        \sum_{i\neq j}
        \log
        \frac{
            P(J_{ij}\mid z_j=+1)
        }{
            P(J_{ij}\mid z_j=-1)
        }
    \right].
    \label{eq:phi_update_general}
\end{equation}
Thus the inferred identity of a neuron combines its prior odds with evidence
accumulated across all of its outgoing couplings.

For the sign-dependent Laplace prior of
Eq.~\eqref{eq:dale_prior}, the two conditional distributions have the
same normalization and their log ratio simplifies to
\begin{equation}
    \log
    \frac{
        P(J_{ij}\mid z_j=+1)
    }{
        P(J_{ij}\mid z_j=-1)
    }
    =
    \beta J_{ij}.
    \label{eq:Dale_log_ratio}
\end{equation}
Substitution into Eq.~\eqref{eq:phi_update_general} gives
\begin{equation}
    \phi_j
    =
    \sigma
    \left[
        \log
        \frac{P(z_j=+1)}{P(z_j=-1)}
        +
        \beta
        \sum_{i\neq j}J_{ij}
    \right],
\end{equation}
which is Eq.~\eqref{eq:phi_update} of the main text.

For fixed $\phi_j$, averaging the conditional log-prior over $Q_j(z_j)$ gives
different effective penalties for positive and negative couplings,
\begin{equation}
    \widetilde\gamma_j^{+}
    =
    \gamma_1+\beta(1-\phi_j),
    \qquad
    \widetilde\gamma_j^{-}
    =
    \gamma_1 + \beta\phi_j,
    \label{eq:Dale_effective_penalties_app}
\end{equation}
which lead directly to the asymmetric proximal update in
Eq.~\eqref{eq:Dale_proximal}.

\subsection{Matched Dale-network experiment}

For the matched-model experiment of Fig.~\ref{fig:DaleAccuracy}, the
ground-truth network contains $80$ excitatory and $20$ inhibitory neurons.
Conditional on the neuronal identity, outgoing couplings are assigned the
corresponding sign, with magnitudes drawn from a Laplace distribution of rate
$\gamma_{\rm gen}=50$. Couplings with magnitude below
$J_{\rm min}=0.015$ are set to zero.

Inference is performed without supplying the true excitatory fraction, using
the uninformative prior
\begin{equation*}
    P(z_j=+1)=P(z_j=-1)=\frac{1}{2}.
\end{equation*}
Inference combines the sparse Laplace prior and its Dale-consistent extension.
We use $\gamma_1=25$ for the baseline sparsity penalty and
$\beta=50$ for the additional sign-inconsistency penalty, while the learning rate is fixed to $\alpha=10^{-6}$. Neuronal
identities are classified according to
\begin{equation*}
    \widehat z_j
    =
    \begin{cases}
        +1, & \phi_j>1/2,\\
        -1, & \phi_j\leq1/2.
    \end{cases}
\end{equation*}
Coupling-sign recovery is evaluated only over ground-truth non-zero
interactions,
\begin{equation}
    A_{\rm sign}
    =
    \frac{
        \sum_{i\neq j}
        \mathbb I[J_{ij}^{\rm true}\neq0]\,
        \mathbb I[
            \operatorname{sign}(J_{ij}^{\rm inf})
            =
            \operatorname{sign}(J_{ij}^{\rm true})
        ]
    }{
        \sum_{i\neq j}
        \mathbb I[J_{ij}^{\rm true}\neq0]
    }.
    \label{eq:Dale_sign_accuracy}
\end{equation}
As in the sparse-network validation, the emission-model parameters are
fixed to their generating values so that the experiment isolates recovery
of the structured interaction matrix.

\section{Conductance-based LIF benchmark}
\label{app:LIF_details}

Here we give the simulation and inference details for the model-mismatch
experiment of Sec.~\ref{sec:LIF}. The benchmark is based on the
conductance-based leaky integrate-and-fire dynamics of
Ref.~\cite{renart2010asynchronous}, with reduced population sizes and modified
synaptic and external-input parameters. We first specify the neuronal and
synaptic dynamics, then the network realization and recording protocol, and
finally the SpiKIsing inference and evaluation procedure.

\subsection{Membrane and synaptic dynamics}

For neuron $i$ in recurrent population $\alpha\in\{E,I\}$, the subthreshold
membrane potential obeys
\begin{equation}
    C_m\frac{dV_i^\alpha}{dt}
    =
    -g_L\left(V_i^\alpha-V_L\right)
    +
    \sum_{\beta\in\{E,I,X\}}
    I_i^{\alpha\beta}(t),
    \label{eq:conductance_LIF}
\end{equation}
where $E$ and $I$ denote the recurrent excitatory and inhibitory populations,
respectively, and $X$ the external input population. No additional applied
current is used.

Synaptic input from population $\beta$ is conductance based,
\begin{equation}
    I_i^{\alpha\beta}(t)
    =
    -
    \left[
        \sum_{j=1}^{N_\beta}
        p_{ij}^{\alpha\beta}
        g_{ij}^{\alpha\beta}
        s_{ij}^{\alpha\beta}(t)
    \right]
    \left(
        V_i^\alpha-V_{\rm rev}^{\beta}
    \right),
    \label{eq:conductance_current}
\end{equation}
where $p_{ij}^{\alpha\beta}\in\{0,1\}$ specifies the presence of a connection
from source $j$ in population $\beta$ to target $i$ in population $\alpha$,
$g_{ij}^{\alpha\beta}\geq0$ is its conductance amplitude, and
$s_{ij}^{\alpha\beta}(t)$ is the corresponding synaptic gating variable.
The reversal potentials are
\begin{equation*}
    V_{\rm rev}^{E}=V_{\rm rev}^{X}=0\,\mathrm{mV},
    \qquad
    V_{\rm rev}^{I}=-80\,\mathrm{mV}.
\end{equation*}

When $V_i^\alpha$ reaches the threshold $V_{\rm th}$, the neuron emits a
spike and its membrane potential is reset to $V_R$ for an absolute refractory
period. We use
\begin{equation*}
    C_m=0.25\,\mathrm{nF},
    \qquad
    g_L=16.7\,\mathrm{nS},
    \qquad
    V_L=-70\,\mathrm{mV},
\end{equation*}
\begin{equation*}
    V_{\rm th}=-50\,\mathrm{mV},
    \qquad
    V_R=-60\,\mathrm{mV}.
\end{equation*}
The absolute refractory periods are $2\,\mathrm{ms}$ and $1\,\mathrm{ms}$
for excitatory and inhibitory neurons, respectively.

Synaptic conductances have finite rise and decay times. Their gating variables
satisfy
\begin{align}
    \tau_d\frac{ds_{ij}^{\alpha\beta}}{dt}
    &=
    x_{ij}^{\alpha\beta}-s_{ij}^{\alpha\beta},
    \nonumber\\
    \tau_r\frac{dx_{ij}^{\alpha\beta}}{dt}
    &=
    \widetilde{\tau}
    \sum_k
    \delta\!\left(
        t-t_{j,k}^{\beta}-d_{ij}^{\alpha\beta}
    \right)
    -
    x_{ij}^{\alpha\beta},
    \label{eq:conductance_synapse}
\end{align}
where $t_{j,k}^{\beta}$ denotes a presynaptic spike time. We take
$\tau_r=1\,\mathrm{ms}$, $\tau_d=5\,\mathrm{ms}$, and
$\widetilde{\tau}=1\,\mathrm{ms}$, with all recurrent and external conduction
delays fixed at $d_{ij}^{\alpha\beta}=1\,\mathrm{ms}$.

\subsection{Network construction and recording protocol}

The recurrent network contains $N_E=30$ excitatory and $N_I=10$ inhibitory
neurons. Each possible off-diagonal recurrent connection is sampled
independently with probability $p=0.2$, while self-connections are excluded.
For a sampled connection from population $\beta$ to population $\alpha$, the
conductance amplitude is drawn according to
\begin{equation}
    g_{ij}^{\alpha\beta}
    =
    \max\left\{
        0,\,
        \overline g^{\alpha\beta}
        \left(
            1+0.5\,\xi_{ij}^{\alpha\beta}
        \right)
    \right\},
    \qquad
    \xi_{ij}^{\alpha\beta}\sim\mathcal N(0,1),
    \label{eq:LIF_conductance_distribution}
\end{equation}
so that negative Gaussian draws are clipped to zero. The nominal conductance
scales (in nano Siemens) are
\begin{align*}
    \overline g^{EE}
    =
    \overline g^{IE}
    &=6\,\mathrm{nS},\\
    \overline g^{EI}
    =
    \overline g^{II}
    &=12\,\mathrm{nS},\\
    \overline g^{EX}
    =
    \overline g^{IX}
    &=30\,\mathrm{nS},
\end{align*}
where the first population index denotes the target and the second the source.

External drive is provided by $N_X=160$ independent Poisson sources, each
firing at $50\,\mathrm{Hz}$. To avoid shared external input in this benchmark,
the external sources are divided into disjoint groups of four, with one group
assigned to each recurrent neuron. No external source therefore projects to
more than one recurrent neuron. Their conductance amplitudes are drawn using
the same clipped-Gaussian prescription as in
Eq.~\eqref{eq:LIF_conductance_distribution}.

The dynamics is simulated with time step $0.05\,\mathrm{ms}$. Between incoming
spike events, the linear synaptic gating equations are propagated using their
exponential solution, and output spike times are recorded at the simulation
resolution. We simulate $100\,\mathrm{s}$ of activity and discard the first
$2\,\mathrm{s}$ as a transient, leaving a $98\,\mathrm{s}$ recording from all
$40$ recurrent neurons.

\subsection{Inference and evaluation}

SpiKIsing inference is performed on the retained $98\,\mathrm{s}$ recording
using the forward-only variational update of Sec.~\ref{sec:Inference} and the
centred coupling parametrization of Appendix~\ref{app:centered-reparam}.
We use a latent interval of
\begin{equation*}
    \Delta t=10\,\mathrm{ms},
\end{equation*}
corresponding to $T=9800$ latent intervals, while the observed spike times
retain their continuous positions within each interval.

The interaction matrix is inferred using the combined sparse and
Dale-structured prior of Sec.~\ref{sec:Dale}. The learning rate chosen was $\alpha=4 \times 10^{-4}$, while the regularization strengths are
\begin{equation*}
    \gamma_1=0.5,
    \qquad
    \beta=2,
\end{equation*}
for the baseline sparsity and additional sign-inconsistency penalties,
respectively. With the
uniform neuronal-identity prior,
\begin{equation*}
    P(z_j=+1)=P(z_j=-1)=\frac12,
\end{equation*}
no information about the true excitatory--inhibitory composition is supplied
to the inference procedure.

The emission parameters are initialized from the spike trains and held
fixed during the first $100$ inference epochs. They are subsequently updated
every five epochs, with the absolute refractory times estimated by the
profile-likelihood procedure of Appendix~\ref{app:DerC}. The interaction
matrix reported in Fig.~\ref{fig:LIF_recovery} is the final iterate after
$200$ epochs.
\newpage
\bibliographystyle{unsrt}
\bibliography{refs}

\end{document}